\documentclass[showpacs,preprintnumbers,amsmath,amssymb,prb,aps,twocolumn]{revtex4}

\usepackage{graphicx}
\usepackage[usenames,dvipsnames]{color} \usepackage{ulem}

 \newcommand{\bq}{{\bf q}} 
 \newcommand{\ep}{{$e$-ph}\:}
\newcommand{\eph}{{$e$-$ph$}\:} \newcommand{\ee}{{$e$-$e$}\:}

\emergencystretch=\maxdimen
\begin{document}
\bibliographystyle{prsty}

\title{Determinant quantum Monte Carlo study of the
two-dimensional single-band Hubbard-Holstein model}

\author{S. Johnston$^{1,2,3}$} 
\author{E. A. Nowadnick$^{3,4}$} 
\author{Y. F. Kung$^{3,4}$} 
\author{B. Moritz$^{3,5,6}$} 
\author{R. T. Scalettar$^{7}$}
\author{T. P. Devereaux$^{3}$} 
\affiliation{$^1$Department of Physics and Astronomy, University of 
    British Columbia, Vancouver, British 
    Columbia, Canada V6T 1Z1} 
\affiliation{$^2$Quantum Matter Institute, University of British 
    Columbia, Vancouver, British Columbia, Canada V6T 1Z4}
\affiliation{$^3$Stanford Institute for Materials and Energy Sciences, 
    SLAC National Accelerator Laboratory and Stanford University, 
    Stanford, CA 94305, USA} 
\affiliation{$^4$Department of Physics, Stanford University, Stanford, 
    CA 94305, USA} 
\affiliation{$^5$Department of Physics and Astrophysics, University 
    of North Dakota, Grand Forks ND 58202, USA} 
\affiliation{$^6$Department of Physics, Northern Illinois University, 
    DeKalb, IL 60115, USA}
\affiliation{$^7$Department of Physics, University of California - 
    Davis, CA 95616, USA}
\date{\today}

\begin{abstract} 
We have performed numerical studies of the Hubbard-Holstein model in two 
dimensions using determinant quantum Monte Carlo (DQMC). 
Here we present details of the method, 
emphasizing the treatment of the lattice degrees of freedom, and then study the    
filling and behavior of the fermion sign as a function of model 
parameters.  We find a region of parameter space with large Holstein coupling 
where the fermion sign recovers despite large values of the Hubbard interaction. 
This indicates that studies of correlated polarons at finite carrier 
concentrations are likely accessible to DQMC simulations.  We then restrict 
ourselves to the half-filled model and examine the evolution of the antiferromagnetic  
structure factor, other metrics for antiferromagnetic and charge-density-wave order, 
and energetics of the electronic and lattice degrees of freedom as a function of 
electron-phonon coupling. From this we find further evidence for a competition  
between charge-density-wave and antiferromagnetic order at half-filling. 
\end{abstract}

 \maketitle 

\section{Introduction} 
The electron-phonon (\eph) interaction is at the heart of 
a number of important phenomena in solids. It can be a dominant 
factor in determining transport properties or produce broken symmetry states 
such as conventional superconductivity\cite{BCS,Migdal1} and/or 
charge-density-wave (CDW) order.\cite{GrunerRMP1988} In systems well described 
by Fermi liquid theory, many of these phenomena are understood 
within the framework of Migdal and Eliashberg theory, which provides a quantitative 
account of this physics.\cite{Migdal1,Migdal2,Eliashberg}  
The situation however can be quite different in correlated systems where the 
role of the \eph interaction is far less well understood, 
sometimes even on a qualitative level. 

From an experimental point of view, interest in the \eph interaction 
in correlated systems has largely been driven by research on transition 
metal oxides, and in particular the high-T$_c$ 
cuprates. 
For example, in undoped Ca$_{2-x}$Na$_x$CuOCl$_2$, angle-resolved 
photoemission spectroscopy (ARPES) studies have found 
broad gaussian spectral features which have been interpreted in terms of 
Franck-Condon processes and polaron physics.\cite{ShenPRL2004} 
This is supported by 
models for a single hole coupled to the lattice and doped into an 
antiferromagnetic background,\cite{BoncaPRB2008,
CataudellaPRL2007,MishchenkoPRL2004}  
which reproduce the observed lineshape and 
dispersion. Similarly, the structure 
of the optical conductivity of the undoped cuprates is well reproduced 
by models with strong (polaronic) \eph coupling.\cite{MishchenkoPRL2008,
DeFilippisPRB2009} These observations point towards 
a strong \eph interaction in the undoped and underdoped cuprates, where 
strong correlations have the largest effect.  

Evidence for lattice coupling also exists in the doped cuprates.  
Perhaps the most discussed are the dispersion 
renormalizations in the nodal 
and anti-nodal regions of the Brillouin zone revealed by 
ARPES.\cite{CukPRL2004,LanzaraNature2001,KordyukPRL2006,JohnsonPRL2001,
VishikPRL2012,PlumbPRL2010,DahmNaturePhysic2009,JohnstonACMP2010}  These manifest as 
sharp changes or ``kinks" in the electronic band dispersion, which 
are generally believed to be due to coupling to a sharp bosonic mode. 
Although the identity of this mode (be it an electronic collective mode 
or one or more phonon modes) remains controversial, the appearance 
of the dispersion renormalizations at multiple energy scales ranging 
from 10 - 110 meV strongly suggests coupling to a spectrum of oxygen 
phonons.\cite{VishikPRL2012,PlumbPRL2010,LeePRB2008,AnzaiPRL2010,
RameauPRB2009,MeevasanaPRL2006}  These electronic renormalizations have 
analogous features in the density of states as probed by  
scanning tunnelling microscopy\cite{LeeNature2006,
Pasupathy,JenkinsPRL2009,deCastroPRL2008,ZasadzinskiPRL2006,
ZhuPRB2006,JohnstonDOS,GuomengZhao} 
as well as in the optical properties of the cuprates.\cite{CarbotteReview,EvH} 

Moving beyond the cuprates, strong \eph and electron-electron (\ee) interactions 
also are believed to be operative in a number of other systems.  These include 
the quasi-1D edge-shared cuprates,\cite{LeePreprint} the manganites,
\cite{MillisNature1998, MillisPRB1996, MannellaPRB2007},  
the fullerenes,\cite{Durand2003,CaponeScience2002,GunnarssonRMP1997,
HanPRL2003}  and the  
rare-earth nickelates.\cite{MedardeJPCM1997, LauPreprint} 
Thus understanding the role of the \eph interaction in 
correlated systems  
is an important problem with possible implications across many 
materials families.   

One of the primary barriers to resolving these issues 
is the incomplete understanding of how 
the direct interplay between the \eph interaction and other important 
degrees of freedom (such as strong \ee 
interactions, magnetic degrees of freedom, reduced dimensionality, 
charge localization, {\it etc}.) influences the \eph interaction.  
On quite general grounds one expects that competition and/or cooperative 
effects can significantly alter the nature of the \ee and \eph interactions.  
Strong \ee interactions will suppress charge fluctuations and will have 
a tendency to localize carriers and renormalize the \eph interaction. 
Conversely, the \eph interaction mediates a retarded attractive 
interaction between electrons that can counteract the
repulsive Coulomb interaction. However, the
interaction with the lattice will further dress quasiparticle mass, 
producing heavier quasiparticles which may be affected more 
significantly by the \ee interaction. In
the limit of strong coupling this can lead to small polaron
formation which also localizes carriers. In the end which, if any, of 
these effects wins out is a complicated question. 

Recent work  
has begun to examine these issues by incorporating the Coulomb interaction 
at varying levels using a variety of analytical and numerical methods.  
This has resulted in a number of interesting results 
which are sometimes contradictory.   
Recent Fermi-liquid based treatments of the 
long-range components of the Coulomb interaction have shown that the \eph 
coupling constant can be significantly enhanced at small momentum transfers 
due to the quasi-2D nature of transport in the cuprates and the breakdown 
of screening in the deeply underdoped samples.
\cite{AlexandrovPRL2011,AlexandrovPRB1996,JohnstonPRB2010,JohnstonPRL2012,MeevasanaScreening,
MeevasanaPRL2006} 
The enhanced coupling in the forward scattering direction can enhance pairing 
in a $d$-wave superconductor\cite{BulutPRB1996} and also affects the energy 
scale of the dispersion renormalization.\cite{MaksimovPRB2005}  
The modification of the \eph vertex appears 
to be generic as studies examining the short-range components of the 
Coulomb interaction as captured by the Hubbard interaction find similar 
forward scattering enhancements of the \eph vertex.\cite{KulicPRB1994,HuangPRB2003,
ZeyherPRB1996}  The short-range Hubbard interaction may also impact 
the energy scale of the \eph renormalizations in the electronic 
dispersion as evidenced by a 
recent dynamical mean-field theory (DMFT) study.\cite{BauerPRB2010} 

Cooperative and competitive effects between the two interactions 
also have been examined in the limit of strong correlations. 
One example of this is in the context of understanding the 
anomalous broadening and softening of the Cu-O bond stretching 
phonon modes in the high-T$_c$ cuprates as a 
function of doping.\cite{Pintschovius,ReznikNature} 
Attempts to account for the observed renormalizations within density functional 
theory have generally been unsuccessful, particularly in the 
case of the phonon linewidth.\cite{DFT,ReznikNature} In contrast, 
correlated multiband and $t$-$J$ models with phonons have experienced  
more success in describing this physics.\cite{HorschPhysicaB2005,RoschPRB2004} 
The most likely origin of this discrepancy is the underestimation of 
correlations and the over-prediction of screening effects within DFT.  

The presence of multiple interactions is also expected to enhance 
quasiparticle masses and therefore influence the formation of small polarons. 
DMFT studies of the Hubbard-Holstein (HH) model have found that the 
Hubbard interaction modifies the critical coupling $\lambda_c$ for the 
crossover to a small polaron.\cite{MishchenkoPRL2004,MacridinPRL2006,BoncaPRB2008,RoschPRB2004} 
However, the suppression or enhancement of $\lambda_c$ depends 
on the underlying phase: paramagnetic (suppression) or antiferromagnetic 
(enhancement).\cite{SangiovanniPRL2005,SangiovanniPRL2006}  
These results indicate the importance of correlations and 
the presence of the underlying magnetic order.   
A Diagrammatic Monte Carlo 
work on the $t$-$J$-Holstein model also found an increased tendency 
towards polaron formation for a single hole doped into an AFM 
background.\cite{MishchenkoPRL2004}  Similar results have  
been obtained in other approaches applied to \eph coupling in $t$-$J$ 
models,\cite{BoncaPRB2008,RoschPRB2004,
RoschPRL2004,MishchenkoPRL2004,Prelovsek} however 
these results are in contrast with 
the exact solution for a two-site HH model where  
$\lambda_c$ increases for increasing Hubbard interaction 
strengths.\cite{BerciuPRB2007} Although this result was obtained 
for a small molecular cluster, it does highlight the need to examine models 
where $U$ is finite in order to allow for the possible destabilization 
of the AFM correlations by the \eph interaction. Without this effect  
it is impossible to address the competition between AFM and a competing order 
driven by the \eph interaction, such as superconductivity or CDWs,  
in an unbiased manner.  

In the case of the HH model the \eph and \ee interactions can 
drive competition between different ordered phases. Take for example the 
half-filled Hubbard and Holstein models on a two dimensional square lattice.  
The single-band Hubbard model has strong ${\bf Q} = (\pi/a,\pi/a)$ 
correlations which favor single occupation of the sites.\cite{WhitePRB1989}  
Conversely, the single-band Holstein model exhibits a ${\bf Q}= (\pi/a,\pi/a)$ CDW phase 
transition at finite temperature.\cite{ScalettarPRB1989,MarsiglioPRB1990} In the 
CDW ordered phase the lattice sites are doubly occupied in a checkerboard pattern.  
When both interactions are present the tendency towards these incompatible orders 
clearly will compete.\cite{BauerPRB2010_2, BauerEPL2010,FehskePRB2004,ClayPRL2005,
NowadnickPRL2012} 
Competing orders in correlated systems is a prominent issue and a common theme 
in many transition metal oxides where novel physics often emerges 
at the boundary between orders.  
 
\begin{figure}
 \includegraphics[width=0.8\columnwidth]{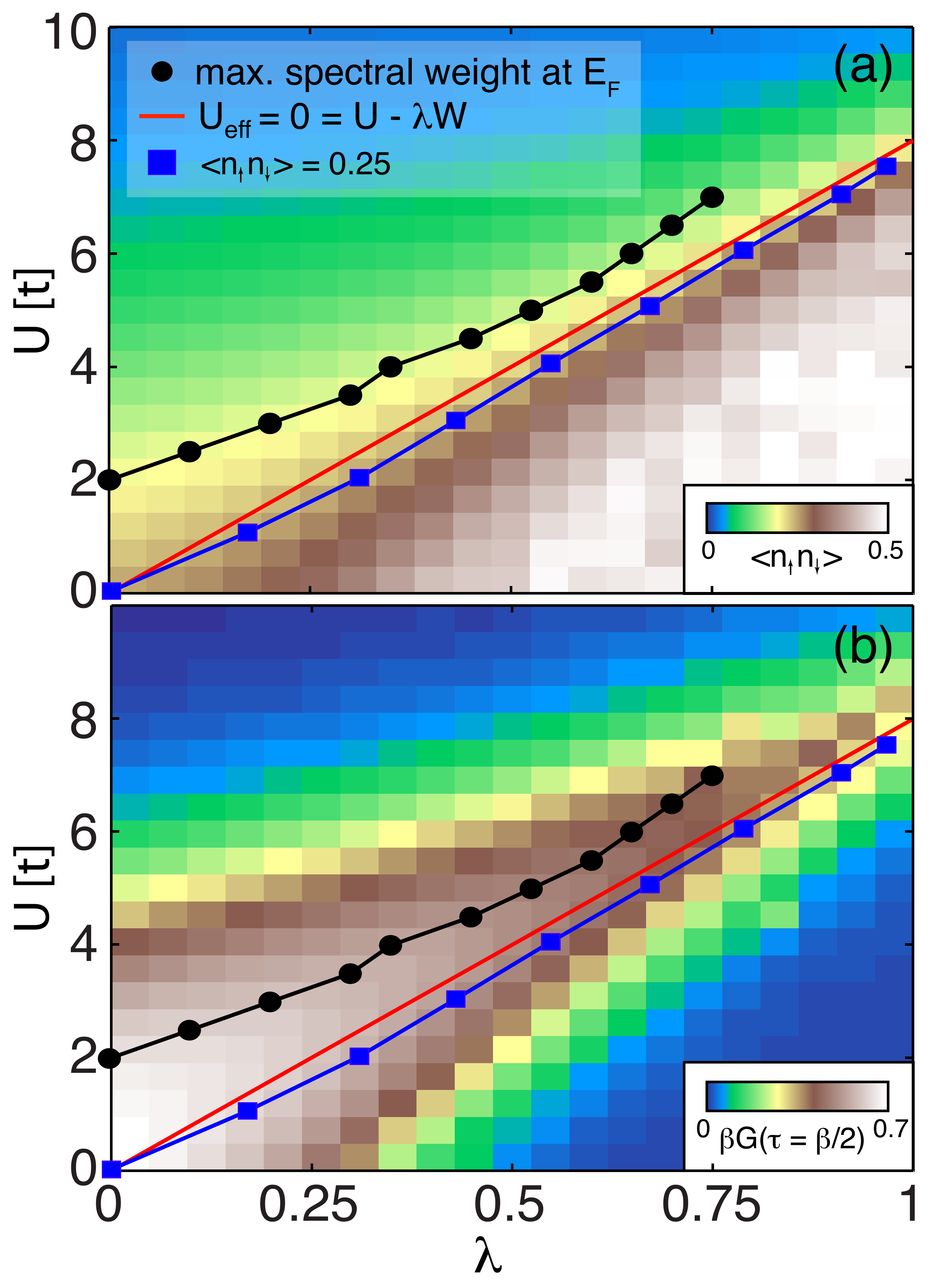}
 \caption{\label{Fig:PD} 
     (color online) The finite temperature ($\beta = 4/t$) 
 phase diagrams for the two-dimensional Hubbard-Holstein model at 
 half-filling.  The vertical 
 axis is the strength of the Hubbard interaction while the horizontal axis 
 is the strength of the Holstein interaction measured in dimensionless units 
 (see text). The color scale in the upper panel gives the average value of 
 the double occupancy per site. In the lower panel it gives the 
 spectral weight at the Fermi surface. For reference, the 
 point of maximum spectral weight is shown in the upper panel and the 
 line where the double occupancy is one quarter is shown in the lower 
 panel. The red line indicates the line where $U_{\mathrm{eff}} = 0$ in 
 the antiadiabatic limit.}
\end{figure}

The $T = 0$ phase diagrams of the half-filled HH model in one and 
infinite dimensions have been mapped 
out.\cite{BauerPRB2010_2, BauerEPL2010,FehskePRB2004,ClayPRL2005} 
Recently this work was extended and a finite temperature  
phase diagram was proposed 
for the two-dimensional (2D) case at half-filling
using determinant quantum Monte Carlo (DQMC).\cite{NowadnickPRL2012}
Fig.~\ref{Fig:PD} sketches the result, extending the 
diagram shown in Fig. 4 of Ref. \onlinecite{NowadnickPRL2012} to include 
additional metrics for the phases involved.  In Fig.~\ref{Fig:PD}a 
the average value of the double occupancy is shown as a function 
of the \ee ($U$) and \eph ($\lambda$, dimensionless units, see below) 
interaction strengths. When the \ee interaction dominates AFM correlations 
develop and $\langle n_\uparrow n_\downarrow\rangle$ is small. 
Conversely, when the \eph interaction dominates $\langle n_\uparrow n_\downarrow\rangle$  
tends towards $0.5$ as half of the sites are doubly occupied 
in a ${\bf Q} = (\pi/a,\pi/a)$ checkerboard pattern.  
These limits are divided by the line where the strength of the \ee interactions 
is comparable to the \eph interaction (indicated by the red line), which is taken 
to be the approximate phase boundary.  

This phase diagram is quite similar to the ones drawn for the one- and infinite-
dimensional cases, however, in the vicinity of the transition there is  
debate as to whether there is an intervening metallic state. 
Here, in the finite $T$ 2D case, we find indications of such a phase.\cite{NowadnickPRL2012} 
This is most clearly seen in the spectral weight at the Fermi level, 
which is related to $G(\tau = \beta/2)$\cite{NowadnickPRL2012} 
and is shown in Fig.~\ref{Fig:PD}b.  To the left (right) of transition region   
spectral weight is suppressed at the Fermi level due to the opening of a Mott (CDW) gap. 
However, in the transition region the spectral weight is maximal, 
consistent with an intervening metallic phase. The point of maximal 
spectral weight lays near the line where 
$\langle n_\uparrow n_\downarrow\rangle = 0.25$, a value equal to that 
expected for a paramagnetic metal. Furthermore, 
as the temperature is lowered, the low energy spectral weight in the 
intervening phase grows, indicative of metallic behavior, 
while the spectral weight in the large $U$ and $\lambda$ regimes 
falls, as expected for an insulator \cite{NowadnickPRL2012}. These results are    
in contrast to the results obtained in infinite dimensions and $T = 0$ 
where a first order AFM/CDW transition has been 
proposed.\cite{BauerPRB2010_2, BauerEPL2010} 
At this stage it is unclear what role dimension and temperature 
are playing, indicating the need for further studies. 



In this paper we apply DQMC to study the 2D single-band HH 
model. 
DQMC is a non-perturbative auxiliary-field technique capable of handling 
both the Hubbard and Holstein interactions on equal footing.  This is particularly 
important if one wishes to address competition between the two interactions 
in an unbiased manner.  
Our results show a number of indications of a competition 
between the ${\bf Q} = (\pi/a,\pi/a)$ CDW and AFM orders.  The primary 
evidence for this has been reported in a previous letter 
(Ref. \onlinecite{NowadnickPRL2012}). 
The purpose of this work is to outline the algorithm, benchmark 
it, and present supporting evidence for the competition between CDW and AFM 
in the half-filled model.  Results are given for 
the fermion sign, which is important for assessing when and  
where it is feasible to apply DQMC.  
For large \ee interactions the fermion sign problem generally restricts 
DQMC simulations to high temperature however, we find a parameter regime
with strong \ee and \ep interactions where the fermion sign recovers.  This 
opens the possibility of treating strongly correlated polarons at finite 
carrier concentrations provided the phonon field sampling remains efficient. 
 
The organization of this work is as follows.  
In the following section we will briefly review the DQMC method as it applies
to the HH model. As previous works\cite{WhitePRB1989,BSS} have
outlined the method in the context of the Hubbard model, here we focus on the
additional aspect associated with the treatment of the lattice degrees of
freedom.  Following this we begin presenting results. 
Section \ref{Sec:sign} examines the severity of the fermion sign problem 
throughout parameter space.  Section \ref{Sec:Filling} examines the filling and 
compressibility of the model as a function of chemical potential. These 
results are intended to provide a reference point for future finite concentration 
studies.  From this point forward we then restrict ourselves to half-filling. 
In section \ref{Sec:Comp} 
we study the AFM structure factor and metrics for the AFM \& CDW orders as 
a function of \eph coupling.  These results provide further evidence of 
the competition between the two orders at half-filling. This competition also is 
evident in the energetics of the electronic and lattice degrees of 
freedom which are presented in section \ref{Sec:energy}. Finally, 
in section \ref{Sec:Discussion} we summarize and make some concluding remarks.

\section{Formalism}\label{Sec:Method} 
In this section we outline the DQMC algorithm.
The general approach follows the original formalism of Refs. \onlinecite{WhitePRB1989}
and  \onlinecite{BSS}.  Here we briefly summarize the method and
highlight the changes and additions required to handle the lattice degrees of freedom.  

\subsection{The Hubbard-Holstein Model}
The HH hamiltonian is a simple
model capturing the physics of itinerant electrons with both \ee and \eph
interactions.  In this model the motion of the lattice sites is described by  
a set of independent harmonic oscillators at each site $i$, with position and momentum
operators $\hat{X}_i$ and $\hat{P}_i$, respectively.  The \ee and \eph
interactions are both treated as local interactions -- the \ee interaction
given by the usual Hubbard interaction while the \eph interaction arises from
the linear coupling of the local density to the atomic displacement
$\hat{X}_i$.  
The HH hamiltonian can be decomposed into $H = H_{el} + H_{lat} +
H_{int}$ where 
\begin{equation} H_{el}=-t\sum_{\langle i,j\rangle,\sigma}
    c_{i,\sigma}^{{\dagger}}c^{\phantom{\dagger}}_{j,\sigma} 
       - \mu\sum_{i,\sigma} \hat{n}_{i,\sigma}, 
\end{equation} 
and 
\begin{equation} H_{lat}=\sum_i
    \left(\frac{M\Omega^2}{2}\hat{X}_i^2 + \frac{1}{2M}\hat{P}^2_i \right), 
\end{equation} 
contain the non-interacting terms for the electron and lattice
degrees of freedom, respectively, and 
\begin{equation}\label{Eq:Hint} H_{int} =
    U\sum_{i}\left(\hat{n}_{i,\uparrow}-\frac{1}{2}\right)
    \left(\hat{n}_{i,\downarrow}-\frac{1}{2}\right) - g\sum_{i,\sigma}
    \hat{n}_{i,\sigma}\hat{X}_i 
\end{equation} 
contains the interaction terms.
Here, $c^\dagger_{i,\sigma}$ ($c_{i,\sigma}^{\phantom{\dagger}}$) creates (annihilates) an
electron of spin $\sigma$ at site $i$, 
$\hat{n}_{i,\sigma} =c^\dagger_{i,\sigma}c^{\phantom{\dagger}}_{i,\sigma}$ 
is the number operator,
$\langle\dots\rangle$ denotes a sum over nearest neighbors, $t$ is the
nearest neighbor hopping, $\Omega$ is the phonon frequency, $U$ and $g$ are
the \ee and \eph interaction strengths, respectively, and $\mu$ is the
chemical potential, adjusted to maintain the desired filling. It
is convenient to define the dimensionless \eph coupling $\lambda =
g^2/(M\Omega^2 W)$, equal to the ratio of the lattice deformation energy
$E_p = g^2/(2M\Omega^2)$ to half the non-interacting bandwidth $W/2 \sim
4t$. 
Throughout this work we use $\lambda$ as a measure of the \eph
coupling strength and set $a = M = t = 1$ as the units of length, mass and
energy, respectively. 

The competition between the Hubbard and Holstein interactions is often 
demonstrated by explicitly integrating out the phonon degrees of 
freedom. After which one obtains an effective dynamic Hubbard 
interaction\cite{BergerPRB1995}
\begin{equation}
U_{\mathrm{eff}}(\omega) = U + \frac{g^2}{M(\omega^2 - \Omega^2)} 
=U - \frac{W\lambda}{1-(\omega/\Omega)^2}. 
\end{equation}
The second term represents the retarded attractive interaction mediated by the 
phonons for $\omega < \Omega$.  In the antiadiabatic 
limit $\Omega \rightarrow \infty$ with $\lambda$ held fixed, 
this interaction becomes instantaneous and 
one is left with an effective Hubbard model with 
$U_{\mathrm{eff}} = U - g^2/M\Omega^2 = U-\lambda W$.  For large values of $\Omega$ 
the behavior the HH model approaches that of the $U_{\mathrm{eff}}$ model. 
However, for small $\Omega$, retardation effects can become important 
as observed in comparisons between the HH and $U_{\mathrm{eff}}$ Hubbard models 
when one examines observables such as the CDW and AFM 
susceptibilities.\cite{NowadnickPRL2012,SangiovanniPRL2005} Nevertheless,   
the frequency-independent $U_{\mathrm{eff}}$ model is used often to describe the 
HH model and recent studies have found that some of the low-energy properties 
of the model can be captured by such an approximation.\cite{SangiovanniPRL2005,
BauerEPL2010}

\subsection{The DQMC Algorithm} 
In general, one wishes to evaluate the finite temperature expectation 
value of an observable $\hat{O}$ given by
\begin{equation}\label{Eq:Trace} 
\langle \hat{O} \rangle = \frac{{\mathrm{Tr}\hat{O}e^{-\beta H}}}{\mathrm{Tr}e^{-\beta H}} 
\end{equation}
where the averaging is performed within the grand canonical ensemble.  In
order to evaluate Eq. (\ref{Eq:Trace}), the imaginary time interval
$[0,\beta]$ is divided into $L$ discrete steps of length $\Delta\tau = \beta/L$. 
The partition function can then be rewritten using the Trotter
formula as \cite{Suzuki} 
\begin{equation}
Z = \mathrm{Tr}(e^{-\Delta\tau L H})
  = \mathrm{Tr}(e^{-\Delta\tau H_{int}}e^{-\Delta\tau K})^L , 
\end{equation} 
where $K$ is the matrix form of the non-interacting terms 
$K = H_{el} + H_{lat}$, and terms of order $tU(\Delta\tau)^2$ and higher have been
neglected.  In many other modern QMC approaches 
this Trotter error is eliminated by using continuous time algorithms.\cite{GullRMP}  
However, with DQMC one has a highly efficient sampling scheme which is 
difficult to implement in a continuous time approach.  We will return to this point 
when we discuss Monte Carlo updates. For our choice of discrete time grids  
the Trotter errors are typically a few percent and difficult to discern against 
the background of statistical errors when evaluating long range correlation and 
structure factors. 

With this discrete imaginary time grid  
the Hubbard interaction terms can now be written in a bilinear form by introducing
a discrete Hubbard-Stratonovich field $s_{i,l} = \pm 1$ at each site $i$ and
time slice $l$. This results in   
\begin{equation}\label{Eq:HS} 
    e^{-\Delta\tau U(\hat{n}_{i,\uparrow}-1/2)(\hat{n}_{i,\downarrow}-1/2)} 
    = A\sum_{s_{i,l} =\\ \pm 1} e^{-\Delta\tau s_{i,l} \alpha
    (\hat{n}_{i,\uparrow}-\hat{n}_{i,\downarrow})} 
\end{equation} 
where $A = \frac{1}{2}e^{-\Delta\tau U/4}$ and 
$\alpha$ is defined by the relation $\cosh(\Delta\tau \alpha) = \exp(\Delta\tau U/2)$
\cite{HS,BSS,WhitePRB1989}.  In the absence of the \eph interaction, the
trace over fermion degrees of freedom can be performed and 
the partition function is expressed as a product
of determinants \cite{BSS} 
\begin{equation} \label{Eq:Z}
 Z = \sum_{s_{i,l}}\det{M_\uparrow}\det{M_\downarrow} 
\end{equation} 
where $M^\sigma = I + B^\sigma_LB^\sigma_{L-1} \dots B^\sigma_1$. 
Here $I$ is an $N\times N$ identity matrix and the $B_l$ matrices are defined as
\begin{equation}\label{Eq:Bmat} B_l^{\uparrow(\downarrow)} =
 e^{\mp\Delta\tau \alpha v(l)}e^{-\Delta\tau K}, 
\end{equation} 
where $v(l)$ is a diagonal matrix whose $i$-th element is the
field value $s_{i,l}$. 
The evaluation of Eq. (\ref{Eq:Z}) now requires a Monte 
Carlo averaging of the auxiliary fields $s_{i,l}$ (see section \ref{Sec:Method}C). 
This expression must be modified when introducing the \eph interaction. 

In order to handle the motion of the lattice, the position operator $\hat{X}_i$
is replaced with a set of continuous variables
$X_{i,l}$ defined on the same discrete imaginary time grid as the
Hubbard-Stratonovich fields.  The momentum operator is replaced with a finite
difference $P_{i,l} = M (X_{i,l+1}-X_{i,l})/\Delta\tau$ and periodic boundary
conditions are enforced on the interval $[0,\beta]$ such that $X_{i,L} =
X_{i,0}$.  In this treatment we recover the proper values for the average
phonon kinetic and potential energy in the non-interacting limit provided the
sampling of the phonon displacements has been done with care.  

With these changes the fermion trace can again be performed and one has  
\begin{equation} 
 Z = \int dX \sum_{s_{i,l}}
 e^{-E_{ph}\Delta\tau}\det{M_\uparrow}\det{M_\downarrow} 
\end{equation}
where $\int dX$ is short hand for integrating over all of the continuous
phonon displacements $X_{i,l}$ and $M^\sigma$ is defined as before 
but with modified matrices 
\begin{equation}
 B_{l}^{\uparrow(\downarrow)} = e^{\mp \Delta\tau \alpha v(l) -
 \Delta\tau gX(l)}e^{-\Delta\tau K}. 
\end{equation} 
The matrix $v(l)$ is defined as before and $X(l)$ is a diagonal 
matrix whose $i$-th diagonal element is $X_{i,l}$.  The factor
$\exp(-E_{ph}\Delta\tau)$ arises from the bare kinetic and potential energy
terms of the lattice Hamiltonian, $H_{lat}$, where 
\begin{equation}
E_{ph} = \frac{M\Omega^2}{2}X^2_{i,l} 
+ \frac{M}{2}\left(\frac{X_{i,l+1}-X_{i,l}}{\Delta\tau}\right)^2.
\end{equation}   
An expression for the numerator of Eq. (\ref{Eq:Trace}) can be obtained in an 
analogous way. 

Most observables can be expressed in terms of the 
single-particle Green's function $G^\sigma(\tau)$. 
For an electron propagating through 
field configurations $\{s_{i,l}\}$, $\{X_{i,l}\}$, the Green's function 
at time $\tau = l\Delta\tau$ is given by\cite{WhitePRB1989}
\begin{eqnarray}\label{Eq:G}
[G^\sigma(l)]_{ij}&=&\langle \hat{T}_\tau c^{\phantom{\dagger}}_{i,\sigma}(\tau) 
    c^\dagger_{j,\sigma}(\tau) \rangle \\ \nonumber 
    &=&[I + B_l^\sigma\dots B_1^\sigma B_L^\sigma\dots B_{l+1}^\sigma]^{-1}_{ij}, 
\end{eqnarray}
where ${\hat{T}_\tau}$ is the time ordering operator. 
The determinant of $M^\sigma$ appearing in Eq. (10) is independent 
of $l$ and is related to the Green's function on any time slice 
$G_\sigma(l)$ by $\mathrm{det}M^\sigma = \mathrm{det}G^{-1}(l)$.  

\subsection{Sampling the Auxiliary fields} 
The sampling of the
Hubbard-Stratonovich and phonon fields is performed using two types of
single-site updates as well as a ``block" update for the phonon fields. In our 
implementation each Monte Carlo step consists of cycling through these  
three types. 

\subsubsection{Hubbard-Stratonovich Field Updates} 
The evaluation of Eq. (\ref{Eq:G}) requires $O(N^3)$ operations.  However, 
once the Green's function $G^\sigma(l)$ is known, the Green's function 
on the next imaginary time slice can be efficiently computed with a set of matrix 
multiplications (an order $O(N^2)$ operation)
\begin{equation}\label{Eq:Wrap}
G^\sigma(l+1) = B^\sigma_{l+1}G^\sigma(l)[B^{\sigma}_{i+1}]^{-1}.
\end{equation}  
This forms the basis for an efficient single site update scheme. 
One begins by computing the Green's function on a single time slice using
Eq. (\ref{Eq:G}). A series of 
updates are then proposed for the Hubbard-Stratonovich fields while holding the
current configuration $\{X_{i,l}\}$ fixed.  This portion follows the
prescription given in Ref. \onlinecite{WhitePRB1989}. One sweeps through all
sites $i$ proposing $s_{i,l}\rightarrow -s_{i,l} =
s_{i,l}^\prime$, which is accepted with probability
\begin{equation}\label{Eq:R} 
R = R^\uparrow R^\downarrow = 
\frac{\mathrm{det}M^{\uparrow\prime}\mathrm{det}M^{\downarrow\prime}}
{\mathrm{det}M^{\uparrow}\mathrm{det}M^{\downarrow}}, 
\end{equation}  
where $M^{\sigma\prime}$ and $M^\sigma$ correspond to the HS
fields with and without the proposed update, respectively.  

Since the 
phonon fields are held fixed during this update, fast Sherman-Morrison 
updates can be performed in the usual manner.\cite{WhitePRB1989}  One has
\begin{equation}
B^\sigma(l) \rightarrow B^{\sigma\prime}(l) = [I + \Delta^\sigma(i,l)]B^\sigma(l) 
\end{equation} 
where the matrix $[\Delta^\sigma(i,l)]_{jk} = \delta_{ik}\delta_{ik}[\exp(\pm 2\Delta\tau 
s_{i,l})-1]$ has a single non-zero element.  The ratio of
determinants can be computed easily from 
\begin{equation} R^\sigma = 1 + 
(1 - [G^\sigma(l)]_{ii})[\Delta^\sigma(i,l)]_{ii}.  
\end{equation} 
If the spin-flip of the Hubbard-Stratonovich field is accepted, the
updated Green's function is given by 
\begin{equation}\label{Eq:SM}
 [G^{\sigma}(l)]^\prime = G^\sigma(l) - 
 \frac{G^\sigma(l)\Delta^\sigma(i,l)[I-G^\sigma(l)]} {1 + [1 -
 G_{ii}^\sigma(l)]\Delta^\sigma_{ii}(i,l)}.  
\end{equation}
$\Delta^\sigma(i,l)$ has a single non-zero element, making 
evaluation of Eq. (\ref{Eq:SM}) straightforward.  
Once updates have been performed for all fields on time slice $l$, $G^\sigma(l)$ 
is advanced to $G^\sigma(l+1)$ using Eq. (\ref{Eq:Wrap}) and the process  
repeated. 

This update scheme is efficient however, it cannot be fully exploited  
in an auxiliary field continuous time approach where one defines time 
slices $\tau_i$ on a variable grid with spacing 
$\Delta\tau_i = \tau_{i+1}-\tau_i$ and sampling is performed over the 
auxiliary fields and number of time slices. 
For a fixed number of time slices  
the methodology outline above holds and 
the fast update scheme can be used.  The difficulty enters when one 
proposes the insertion or removal of a time slice from the set.  These updates are 
accepted with a probability related to the ratio of  
determinants similar to Eq. (\ref{Eq:R}) times an additional prefactor 
to satisfy detailed balance.\cite{GullRMP}  However, the new configuration 
in this case involves a different number of time slices and thus 
the determinants must be computed from scratch, which is computationally 
expensive.  Since continuous time approaches require many of these 
types of updates we choose to remain on a discrete grid where fast sampling 
of the auxiliary fields can be maintained on larger clusters.   

\subsubsection{Phonon Field Updates} Single-site updates for the phonon
fields proceed in a manner analogous to that for the Hubbard-Stratonovich fields.  
For each point $(i,l)$ one proposes
updates $X_{i,l} \rightarrow X^\prime_{i,l} = X_{i,l} + \Delta X_{i,l}$ while
holding the configuration $\{s_{i,l}\}$ fixed.  In this case $\Delta X_{i,l}$ is
drawn from a box probability distribution function.
The proposed phonon update is then accepted with probability $R =
R^\uparrow R^\downarrow \exp(-\Delta\tau\Delta E_{ph})$ where $\Delta E_{ph}$ is the
total change in kinetic and potential energy associated with the update, and
$R^\sigma$ is defined by Eq. (\ref{Eq:R}).  The $\Delta E_{ph}$ term accounts for
the contribution of $H_{lat}$ to the total action.  The fast Sherman-Morrison
update scheme can also be performed for single-site phonon updates with
$\Delta^\sigma(i,l)$ replaced by 
\begin{equation} 
    [\Delta^\sigma(i,l)]_{jk} = \delta_{ik}\delta_{jk}
    [\exp(-\Delta\tau \Delta X_{i,l})-1].  
\end{equation}

\subsubsection{Block Updates for the Phonon Fields} As noted previously,
sampling the phonon fields requires some additional care.  In addition to the
single-site update scheme we have found that a {\it block
update} scheme is necessary to reproduce correct 
results in the non-interacting and atomic limits.  
In this update scheme the lattice position for a given
site is updated such that $X_{i,l} \rightarrow X_{i,l} + \Delta X$ for all $l
\in [0,L]$.\cite{ScalettarPRB1991} 
This type of update helps to efficiently move the phonon 
configurations out of false minima at lower temperatures.  
However, it comes at a price.  Block updates spanning multiple imaginary time slices are
computationally expensive within the DQMC formalism.  They require that the  
Green's function be recalculated from scratch since updates are being made on 
multiple time slices simultaneously.   
This is an $O(N^3)$ operation in contrast 
to the $O(N^2)$ cost of Eq. (\ref{Eq:SM}).  Therefore a balance between the two
types of phonon updates must be struck. 
As a rule of thumb we have found that
two to four block updates at randomly selected sites for every full set of single
site updates to $\{s_{i,l}\}$ and $\{X_{i,l}\}$ is sufficient to recover the correct
behavior in the non-interacting and atomic limits. 
In our
implementation $\Delta X$ is drawn from a separate box 
probability distribution function.   

\section{The Fermion Sign}\label{Sec:sign} 
\begin{figure}[tl] \includegraphics[width=0.9\columnwidth]{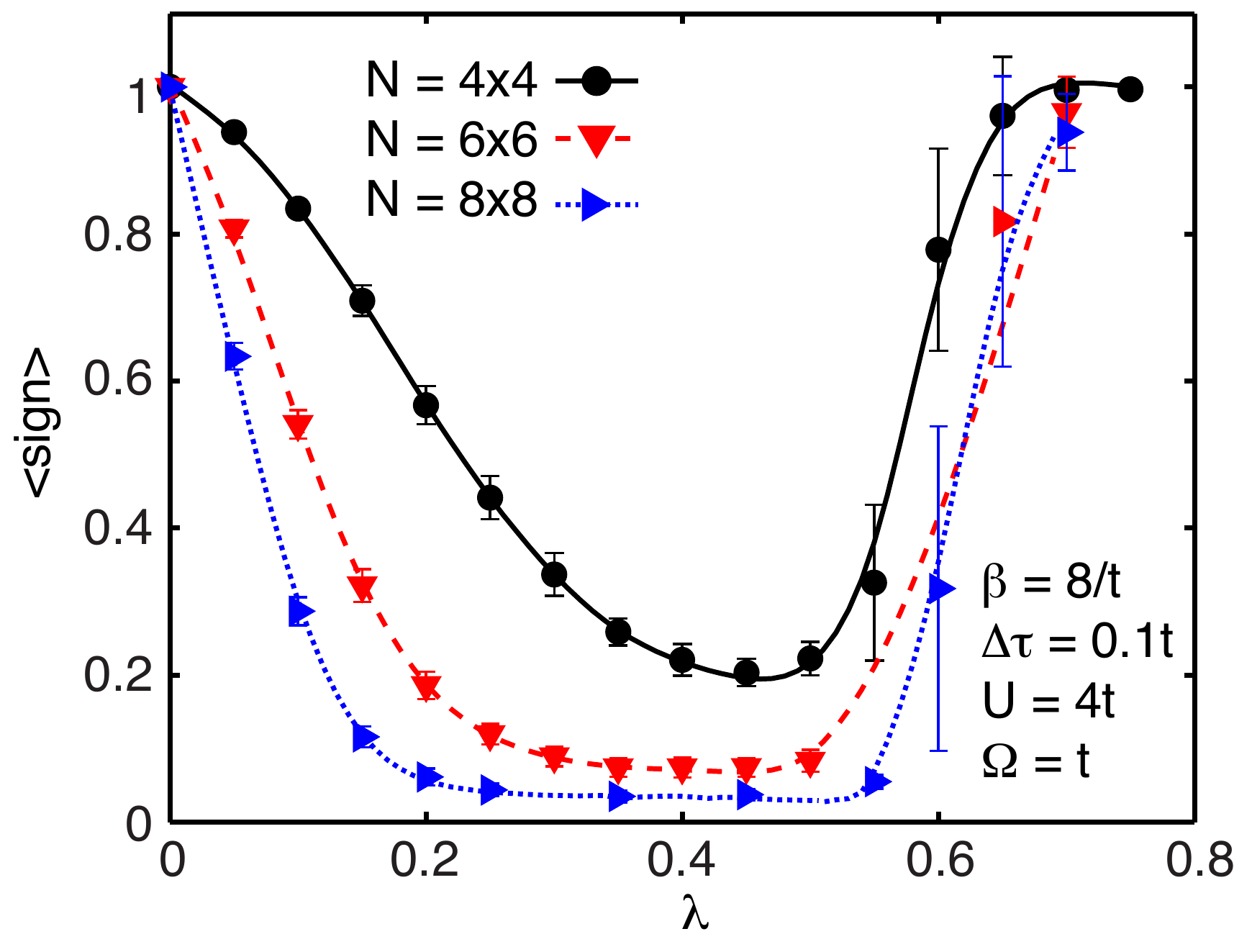}
    \caption{\label{Fig:Sgn_vs_lambda} (color online) The average value of the
        fermion sign as a function of \eph coupling $\lambda$ for various
        half-filled clusters.  The parameters for these calculations are $\beta
    = 8/t$, $\Delta\tau = t/10$, $U = 4t$, and $\Omega = t$}
\end{figure}

We begin with the average value of the fermion sign, which is the limiting factor
for any QMC treatment of correlated electrons.  In Fig.~\ref{Fig:Sgn_vs_lambda}
we focus on the average sign at half-filling as a function of \ep coupling for
a moderately correlated case ($U = 4t$).  Results are shown for a phonon
frequency $\Omega = t$ and inverse temperature $\beta = 8/t$.  Since we
have only included nearest neighbour hopping, the average sign at half-filling
is protected by particle-hole symmetry for $\lambda = 0$.\cite{WhitePRB1989}
This protection results from the fact that although $\mathrm{det}M_\sigma < 0$, 
symmetry dictates $\mathrm{sign}(\mathrm{det}M_\uparrow) = 
\mathrm{sign}(\mathrm{det}M_\downarrow)$ in a particle-hole symmetric system; 
thus the ratio $R$ remains positive definite. 
This no longer holds for finite \eph coupling 
since most phonon configurations $\{X_{i,l}\}$ break this 
symmetry leading to a sign problem at half-filling.  
Increasing $\lambda$ suppresses the average sign until reaching a minimum
that depends on the cluster size.  For larger clusters this
minimum persists over a wide range of $\lambda$; however, the average sign
eventually recovers in all cases when $W\lambda \gtrsim U$.  
This behavior is generic
for all parameter sets we have examined at half-filling and is 
due to the strong reduction of $U_{\mathrm{eff}}$ produced by the
attractive interaction mediated by the \eph interaction.  This result indicates
that although simulations of the HH model at low $T$ remain limited by the
fermion sign problem for arbitrary parameter ranges, this need not be true for 
simulations of the correlated polaronic regime (large $\lambda$ with 
moderate to large $U$).  

\begin{figure}[tr]
    \includegraphics[width=0.75\columnwidth]{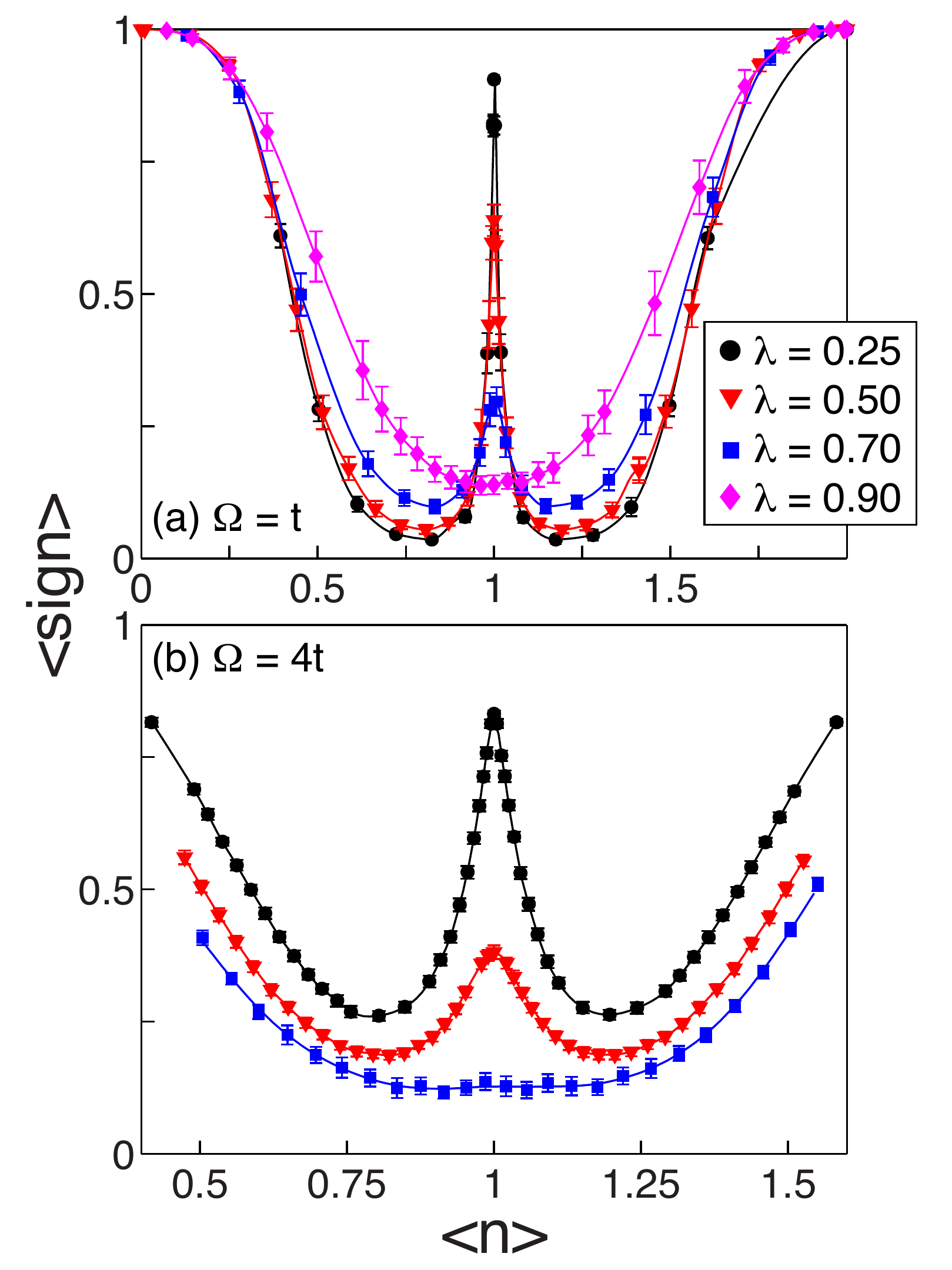}
    \caption{\label{Fig:Sgn} (color online) The average value of the fermion
        sign as a function of filling for $\lambda = 0.25$ ($\circ$),  $0.5$
        ($\triangledown$), $0.7$ ($\square$), and $0.9$ ($\diamond$). Results
        are shown for two sets of phonon frequencies $\Omega = t$ (panel a) and
        $\Omega = 4t$ (panel b).  All data sets are for the strongly correlated
        limit with $U = 8t$.  These results were obtained on a $N = 8\times 8$
    cluster with $\Delta\tau = 0.1/t$.  The inverse temperatures for panels (a)
and (b) are $\beta = 4/t$ and $3/t$, respectively.  The solid lines are guides
to the eye.  } \end{figure}

Turning to finite carrier concentrations, Fig.~\ref{Fig:Sgn} shows the
average sign as a function of filling for a strongly
correlated system ($U = 8t$), phonon frequencies $\Omega = t$ (Fig.~\ref{Fig:Sgn}a) and
$\Omega = 4t$ (Fig.~\ref{Fig:Sgn}b) (the latter being closer to the antiadiabatic
limit), and inverse temperatures are $\beta = 4/t$ and $3/t$,
respectively.  For weak \ep coupling doping suppresses the sign in a
manner similar to the bare Hubbard model \cite{WhitePRB1989} 
where the most severe sign problem occurs near $\langle n \rangle \sim 0.85$ 
and $\sim 1.15$.  
Upon increasing $\lambda$, the behavior at half-filling follows that 
shown in Fig.~\ref{Fig:Sgn_vs_lambda}. However, at finite doping,  
the evolution of the fermion sign depends on the phonon frequency.
For $\Omega = t$ the average value of the sign increases with the 
inclusion of the \eph interaction for most
carrier concentrations away from the immediate vicinity of half filling.  
Conversely, for $\Omega = 4t$, the average sign is systematically suppressed and a
deep minimum develops over a wide doping range for the largest values 
of $\lambda$ considered. This indicates that the way in which the \eph coupling 
affects the sign problem depends both on the strength of the effective attraction 
as well as retardation effects.  We will return to this point shortly. 
Fig.~\ref{Fig:Sgn} also shows that for large 
$\lambda$, the degree to which the sign is enhanced or suppressed  
at finite doping is comparatively smaller than the size of the 
induced sign problem at half filling.  In other words, although
a sign problem is induced at half-filling, it does not appear to be
significantly exacerbated, and can even be improved by the \eph interaction, 
near carrier concentrations that are of interest for the doped 
high-T$_c$ cuprates.  

\begin{figure} 
    \includegraphics[width=0.8\columnwidth]{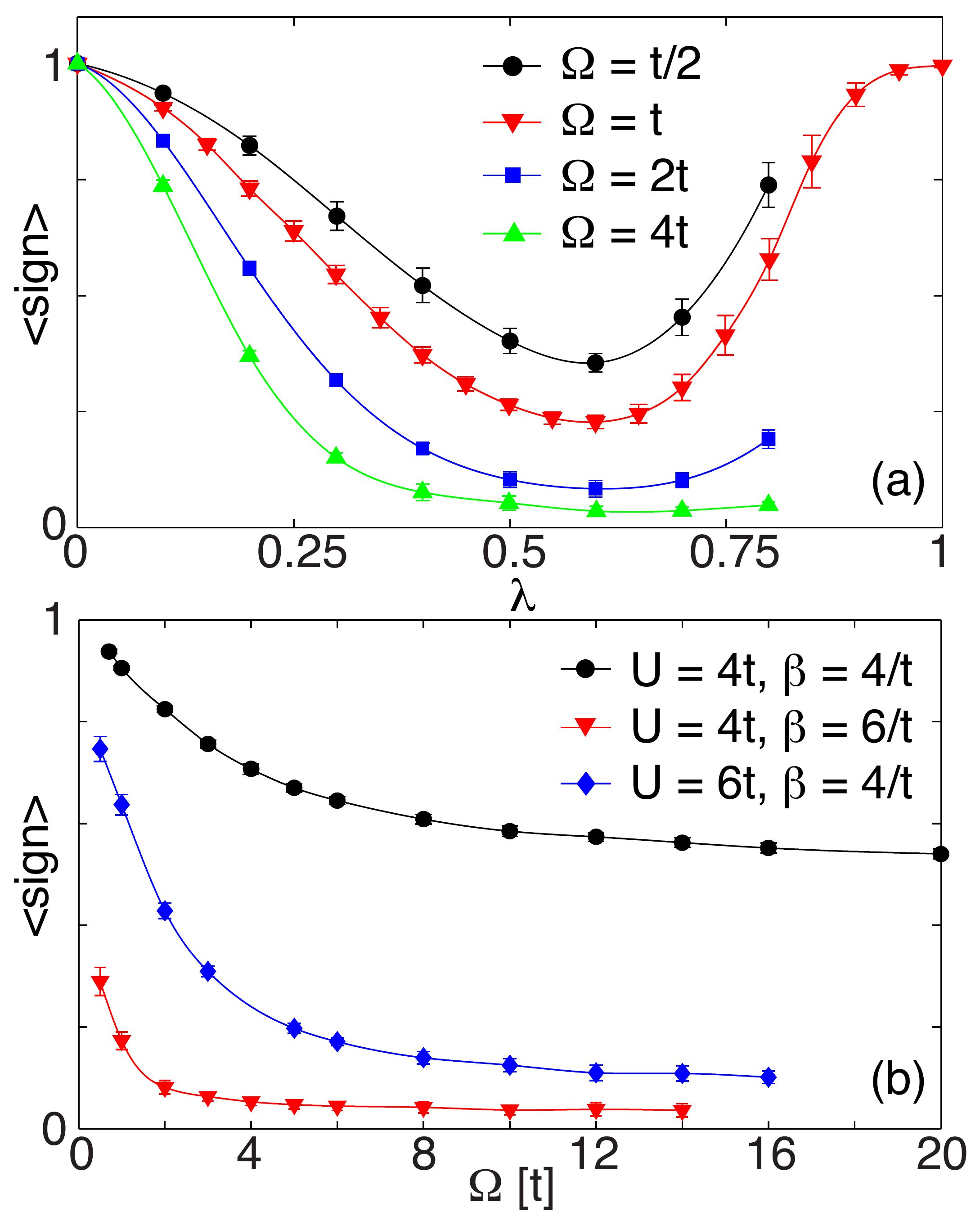}
    \caption{\label{Fig:sgn_vs_omega} 
        (color online) (a) The average sign for
        $\langle n \rangle = 1$ as a function of $\lambda$ for $\Omega = t/2$
        ($\circ$), $t$ ($\triangledown$), $2t$, ($\square$), and $4t$
        ($\vartriangle$).  The Hubbard interaction strength is 
        held fixed at $U = 6/t$, and the inverse temperature is $\beta = 4/t$.  
        (b) The average value of the fermion sign at half-filling as a function of the 
        phonon frequency $\Omega$ and fixed \eph coupling $\lambda = 0.25$. 
        Results are shown for $U = 4t$,$\beta = 4/t$ ($\circ$), 
        $U = 4t$, $\beta = 6/t$ ($\triangledown$), and $U = 6t$, 
        $\beta = 4/t$ ($\Diamond$). All results in panels (a) and (b) were   
        obtained on an $N = 8\times 8$ clusters with $\Delta\tau = 0.1/t$.   
} \end{figure}

The $\Omega$ dependence of the average sign reinforces the notion that the
degree of retardation associated with the \eph interaction plays an important
role in determining the dressing of the Hubbard interaction.  To explore this
further in Fig.~\ref{Fig:sgn_vs_omega}a we show the average sign at
half-filling for $t/2 < \Omega < 4t$ as a function of $\lambda$.  For a given
value of $\Omega$  the overall trend remains similar to Fig.~\ref{Fig:Sgn_vs_lambda}, 
however, increasing $\Omega$ results in a greater overall suppression of the
average sign, indicating that $U$ is suppressed more rapidly 
by antiadiabatic phonons.  The opposite trend was observed in the AFM 
susceptibilities, where AFM was suppressed at lower values of 
$\lambda$ for   
larger $\Omega$.\cite{NowadnickPRL2012} This suggests that the fermion sign 
is influenced both by the magnitude of $U$ and the degree of retardation 
encoded in $U_{\mathrm{eff}}(\omega)$.\cite{auto} This possibility is underscored by 
contrasting the instantaneous $U_{\mathrm{eff}}$ model to the HH model with 
large $\Omega$. In the $U_{\mathrm{eff}}$ model particle-hole symmetry holds and the 
average 
value of the sign is identically one.  In contrast, we observe that the 
sign is lower for $\Omega$ approaching the antiadiabatic limit as 
shown in Fig. \ref{Fig:sgn_vs_omega}b for a fixed $\lambda = 1/4$. 
Futhermore, the average sign is suppressed more rapidly for small 
$\Omega$ before asymptotically approaching a $U$- and $\beta$-dependent 
value at high frequency. We interpret the value of the sign  
at large $\Omega$ as the size of the induced sign problem introduced 
by the breaking of particle-hole symmetry by the phonon fields. 
A possible explanation for the improved sign at small 
$\Omega$ is the attractive \eph-mediated interaction for 
electrons at the Fermi level.  Recall that the dynamic effective Hubbard 
interaction introduced by the phonons $U^{ph}_{\mathrm{eff}}(\omega)$ 
is attractive for $\omega < \Omega$ and, 
and divergent for $\omega \rightarrow \Omega$. Thus as the phonon frequency 
tends to smaller values, a significant suppression of the repulsive 
Hubbard interaction occurs for electrons in a window near the Fermi level.  
If the average sign is determined primarily by electrons in this window 
then one would expect the sign to be improved.  Further work is clearly 
needed to clarify this interesting possibility. 

\section{Filling and Compressibility}\label{Sec:Filling}
\begin{figure}[tr]
    \includegraphics[width=0.75\columnwidth]{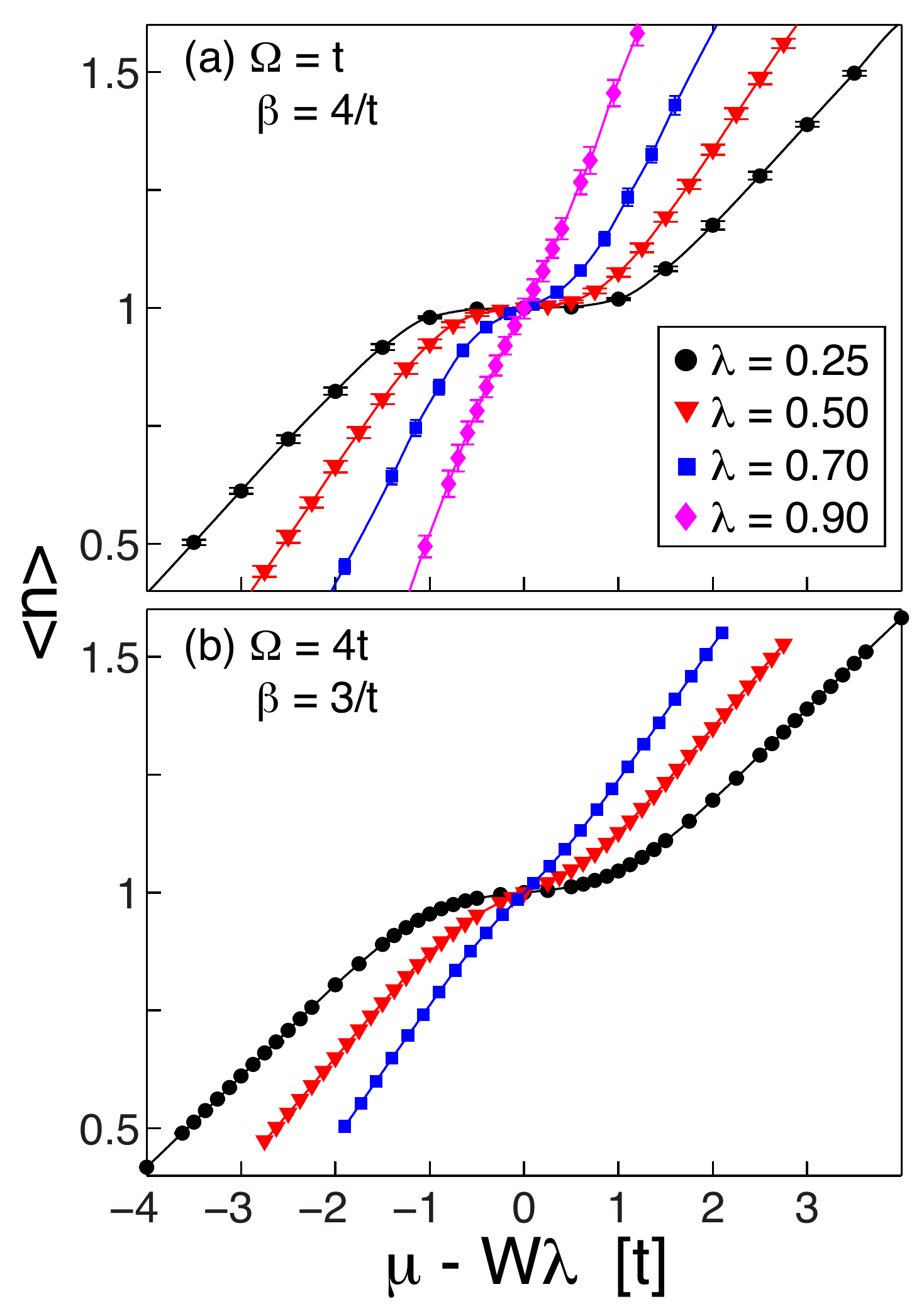}
    \caption{\label{Fig:filling} (color online) 
    The average value of the filling $\langle n \rangle$ as a function of 
    chemical potential $\mu - W\lambda$ for the same parameter sets shown in 
    Fig. \ref{Fig:Sgn}.  The $-W\lambda$ correction accounts for the 
    global shift of the lattice equilibrium position (see main text). 
    }
\end{figure}

Fig. \ref{Fig:filling} shows the average filling on an $8\times 8$ cluster  as a
function of chemical potential $\mu$ for the same parameter set used 
to obtain the results shown in Fig.
\ref{Fig:Sgn}. (A chemical potential shift $\Delta\mu = -W\lambda$ 
due to the equilibrium lattice position 
has been subtracted off such that $\mu =- 0$ corresponds to half-filling, 
see appendix \ref{Sec:lattice}.)  
Fig. \ref{Fig:Compressibility} shows the corresponding compressibility 
$\kappa \propto \frac{\partial \langle n\rangle}{\partial\mu}$ 
for the system.\cite{FootNote1}  In these results 
one starts to see indications of competition between the 
attractive interaction mediated by
the \ep interaction and the repulsive \ee interaction.  
For small values of 
$\lambda$ the strong Hubbard interaction ($U = 8t$) dominates, opening a Mott
gap in the system which clearly manifests as a plateau in $\langle
n(\mu)\rangle$ and incompressibility $\kappa \sim 0$ located 
near $\mu - W\lambda = 0$.
As the strength of the \ep interaction increases the 
effective attractive interaction grows.  This reduces the influence of the
Hubbard interaction and the size of the Mott gap begins to 
diminish. This is evident in the shrinking width of the plateau in 
$\langle n(\mu)\rangle$ and the rise in the value of $\kappa$.    
In the limit of large $\lambda$ all indications of the Mott gap vanish 
and $\langle n(\mu)\rangle$ behaves in a manner 
expected for a metallic state. The system has a   
finite compressibility and $\kappa \rightarrow 0$ as the band completely 
fills.  
This qualitative behavior occurs for both 
phonon frequencies and is further evidence for the direct competition between 
the attractive \eph interaction and repulsive \ee interaction discussed in 
Ref. \onlinecite{NowadnickPRL2012}.  For this parameter set,  
$\lambda \sim 1$ marks the position where one expects the transition 
between the AFM and CDW order (see Fig. \ref{Fig:PD}). 
We interpret this as further evidence for an intervening 
metallic state between the two orders at finite temperature.  
Finally, for the largest coupling $\lambda = 0.9$ 
the $\kappa \rightarrow 0$ for $\mu-W\lambda \rightarrow 3t$ indicating 
that the total bandwidth of the system has been narrowed by the interactions 
present in the system.  

\begin{figure}
    \includegraphics[width=0.75\columnwidth]{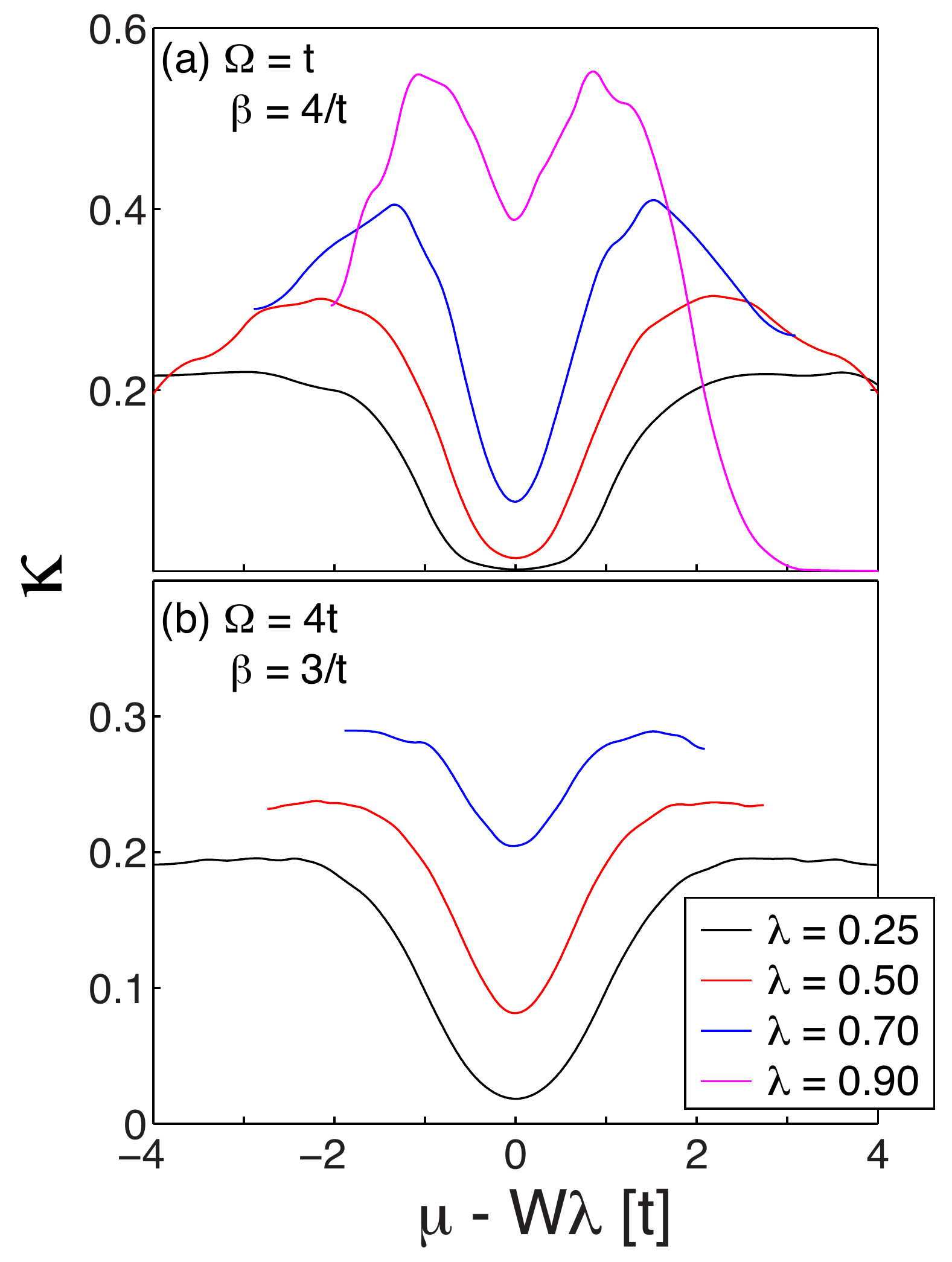}
    \caption{\label{Fig:Compressibility} (color online) 
    The compressibility $\kappa$ as a function of 
    chemical potential for the same parameter set 
    shown in Fig. \ref{Fig:filling}.\cite{FootNote1}}
\end{figure}

\section{Charge-density-wave and Antiferromagnetic Correlations}\label{Sec:Comp} 
\begin{figure}
 \includegraphics[width=\columnwidth]{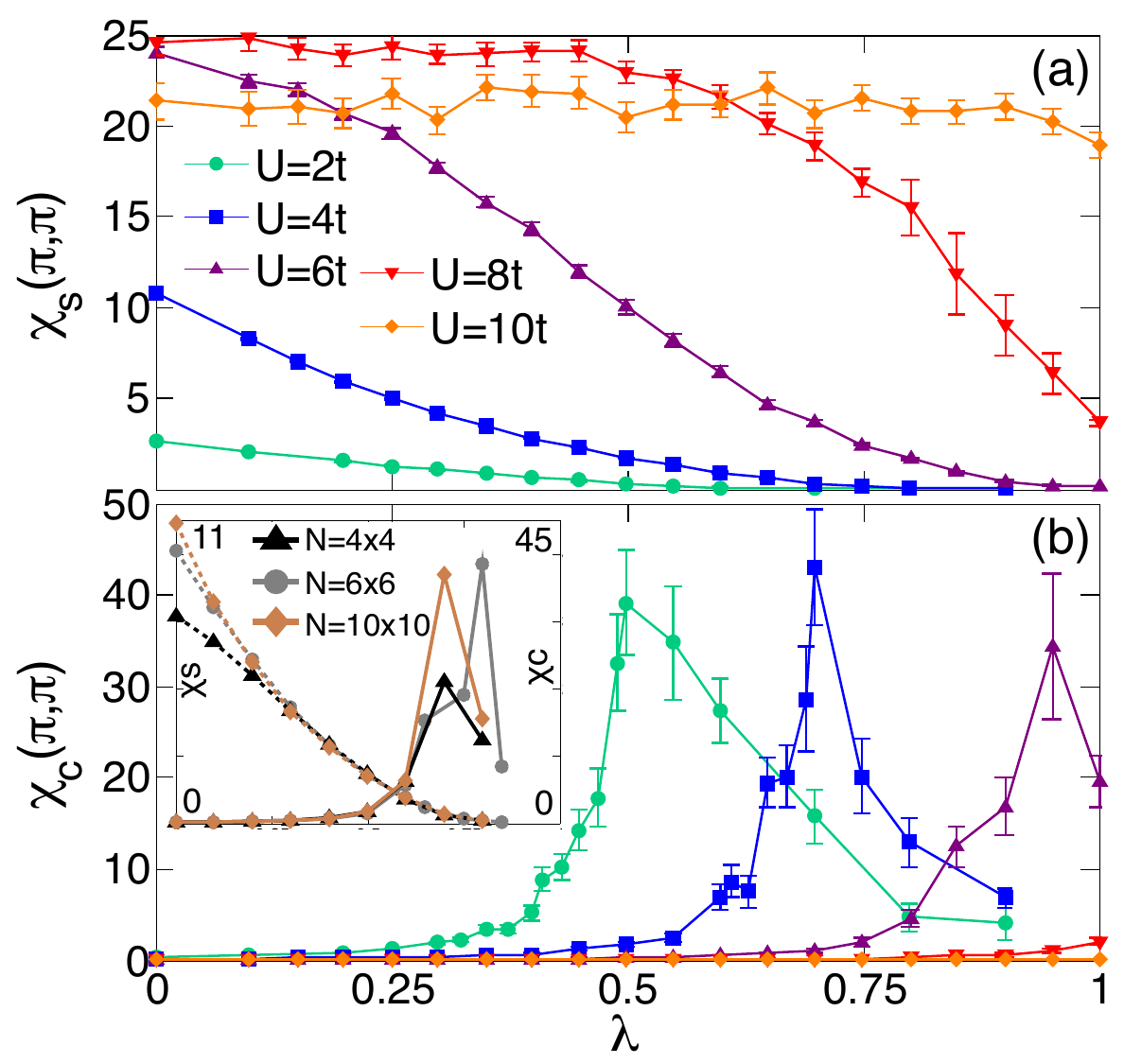}
 \caption{\label{Fig:Susceptibilities} (color online). 
 The (a) spin $\chi_s(\pi,\pi)$ and (b) charge $\chi_c(\pi,\pi)$ 
 susceptibilities for several values of U on an $N = 8\times 8$ cluster, 
 reproduced from Ref. \onlinecite{NowadnickPRL2012}.
 The inset of (b) shows $\chi_s$ (dashed lines) and $\chi_c$ (solid lines) 
 at $U = 4t$ for several lattice sizes.  The error bars in the 
 inset have been suppressed for clarity. The remaining parameters 
 are $\beta = 4/t$, $\Delta\tau = 0.1/t$, and $\Omega = t$.   
} 
\end{figure}

In this section we address the issue of competition between the \eph-driven CDW 
and \ee-driven AFM correlations for the model at half-filling.  We begin by 
first reviewing our previous results 
for the charge $\chi_c(\bq)$ and spin $\chi_s(\bq)$ susceptibilities, 
defined as  
\begin{equation}
\chi_{s,c}(\bq) = \frac{1}{N}\int_0^\beta d\tau 
\langle T_\tau \hat{O}^{\phantom{\dagger}}_{s,c}(\bq,\tau)
\hat{O}_{s,c}^\dagger(\bq,0)\rangle,
\end{equation}
where 
$\hat{O}_{s}(\bq) = \sum_{i} e^{i\bq\cdot{\bf R_i}}(\hat{n}_{i,\uparrow} - 
\hat{n}_{i,\downarrow})$, and 
$\hat{O}_c(\bq) = \sum_{i,\sigma}e^{i\bq\cdot{\bf R_i}}\hat{n}_{i,\sigma}$. 

Our results for $\chi_{s}(\pi,\pi)$ and $\chi_c(\pi,\pi)$  
are reproduced in Fig. \ref{Fig:Susceptibilities}   
as a function of $\lambda$ and for several values of $U$.\cite{NowadnickPRL2012} 
For increasing \eph coupling, 
$\chi_s$ (Fig. \ref{Fig:Susceptibilities}a) is suppressed as a 
result of the reduction in the effective Hubbard interaction.  
For small values of $U$, $\chi_s$ is suppressed  
immediately for finite $\lambda$.  However, for larger values of $U$, 
where more robust AFM correlations are present, $\chi_s$ persists 
up to $\lambda \sim U/W$ before beginning a significant drop 
as a function of $\lambda$. (This is seen most clearly in the data 
for $U = 8t$.) At the same time, as $\lambda$ increases    
there is a corresponding increase in $\chi_c(\pi,\pi)$ (Fig. \ref{Fig:Susceptibilities}b).  
This occurs gradually at first    
while $\chi_s$ is large, but once the AFM correlations have been 
suppressed sufficiently there is a sharp increase in the 
growth of $\chi_c$. This indicates a competition between the 
two orders as the AFM correlations must be suppressed before charge ordering  
can occur.   
Finally, for $U \le 6t$, further increases in $\lambda$ result in a decreasing 
$\chi_c$.  We interpret this as being due to the finite CDW transition 
temperature in the HH model.\cite{NowadnickPRL2012}  
The inset of Fig. \ref{Fig:Susceptibilities} shows similar results 
obtained on different lattices, demonstrating that the finite size 
effects do not qualitatively alter this picture.   

Another measure of the AFM correlations in the single-band model 
can be obtained from the magnitude of the equal-time spin structure 
factor $S(\pi,\pi)$, which is defined as the Fourier transform of the 
spin-spin correlation function $c(l_x,l_y)$ \cite{WhitePRB1989}
\begin{equation}\label{Eq:Sq}
    S(\bq) = \sum_{\bf l} e^{i\bq\cdot{\bf l}} c(l_x,l_y)
\end{equation}
where ${\bf l} = (l_x,l_y)$ is the lattice position and 
\begin{equation*}
c(l_x,l_y) = \frac{1}{N}\sum_{{\bf i}} 
\langle (\hat{n}_{{\bf i}+{\bf l},\uparrow}-\hat{n}_{{\bf i}+{\bf l},\downarrow})
(\hat{n}_{{\bf i},\uparrow}-\hat{n}_{{\bf i},\downarrow})\rangle. 
\end{equation*}
Here the sum over ${\bf i}$ has been introduced to average over translationally 
equivalent quantities as opposed to a non-trivial spatial sum 
as in Eq. (\ref{Eq:Sq}).  

\begin{figure}
    \includegraphics[width=0.75\columnwidth]{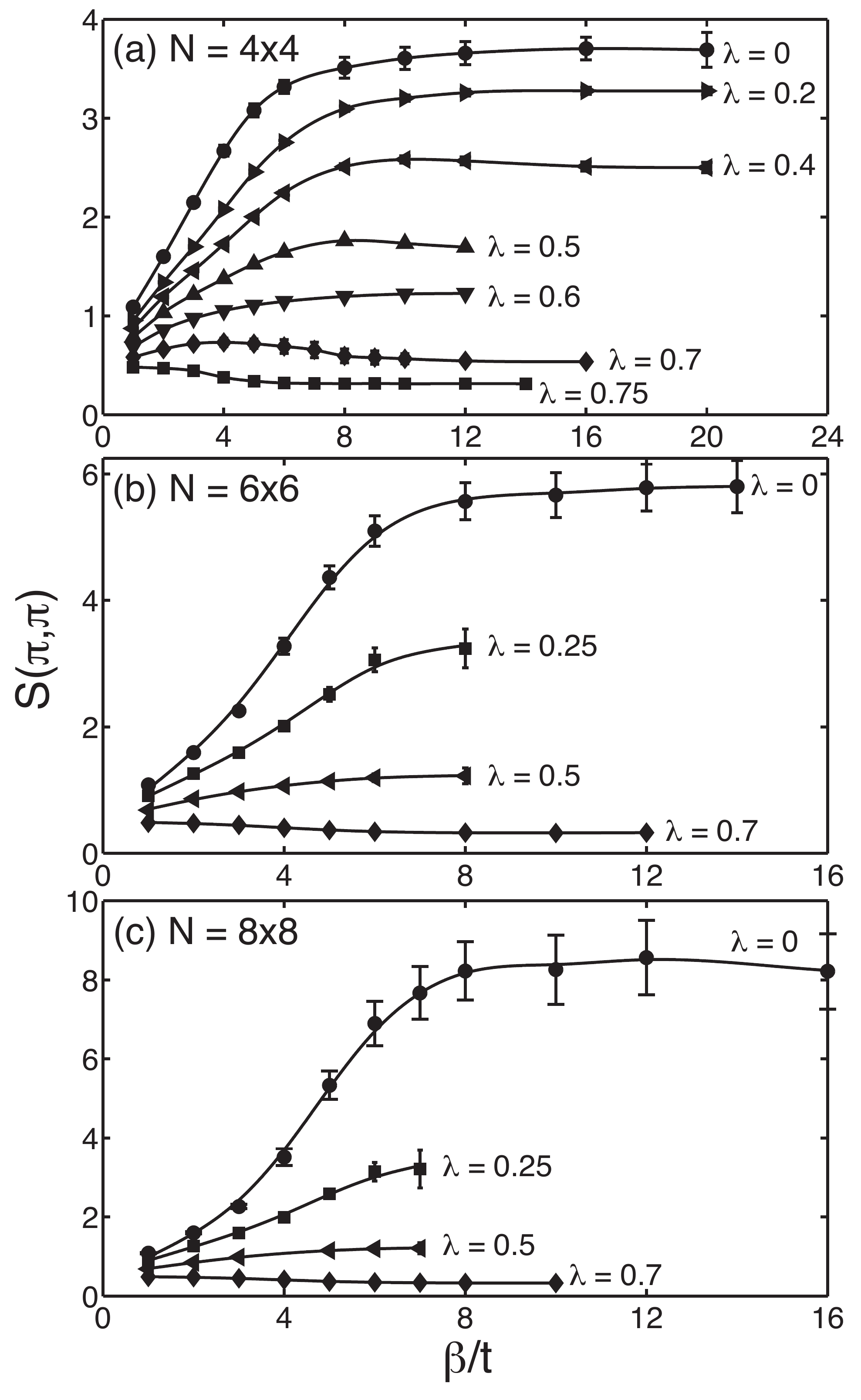}
    \caption{\label{Fig:Spipi} The structure factor 
    $S(\pi,\pi)$ as a function of inverse temperature $\beta$ 
    for the half-filled Hubbard-Holstein model.  Results 
    are shown for clusters of linear dimension (a) $N = 4$, (b) $N = 6$, and 
    (c) $N = 8$ and for several values of the electron-phonon 
    coupling strength $\lambda$, as indicated.  The remaining parameters 
    are $U = 4t$, $\Delta\tau = 0.1t$, and $\Omega = t$.  
}
\end{figure}

\begin{figure}
    \includegraphics[width=\columnwidth]{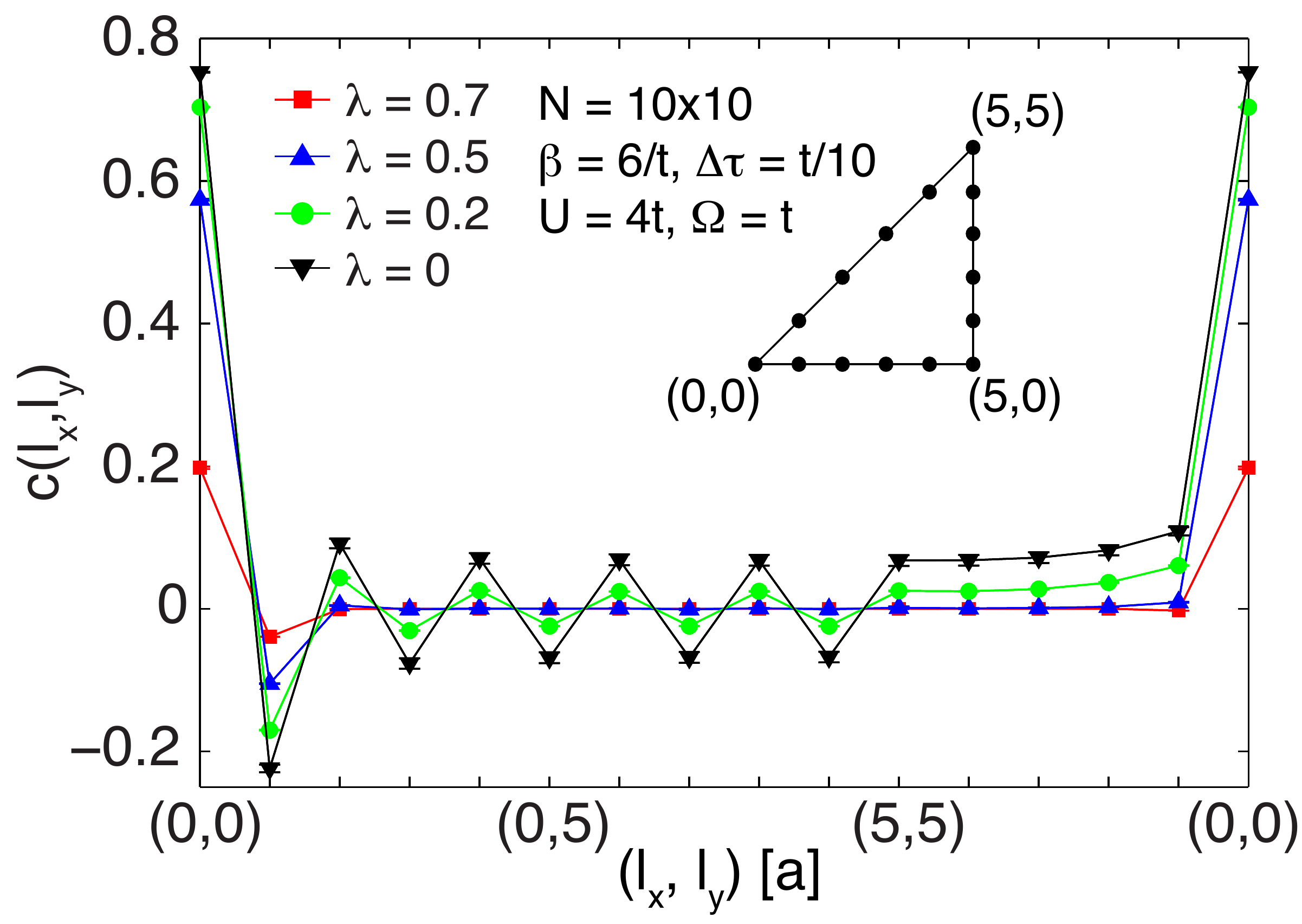}
    \caption{\label{Fig:SCF} (color online) 
    The real space structure of the 
    spin-spin correlation function $c(l_x,l_y)$ along the path indicated in the 
    inset.  For $\lambda = 0$ (black $\triangledown$) the antiferromagnetic 
    correlations are evident. For increasing values of $\lambda$ the antiferromagnetism 
    is suppressed.  By $\lambda = 0.5$ 
    (blue $\triangle$), where $\lambda W \sim U$, all traces of antiferromagnetic 
    correlations are gone. 
    }
\end{figure}

\begin{figure}
    \includegraphics[width=\columnwidth]{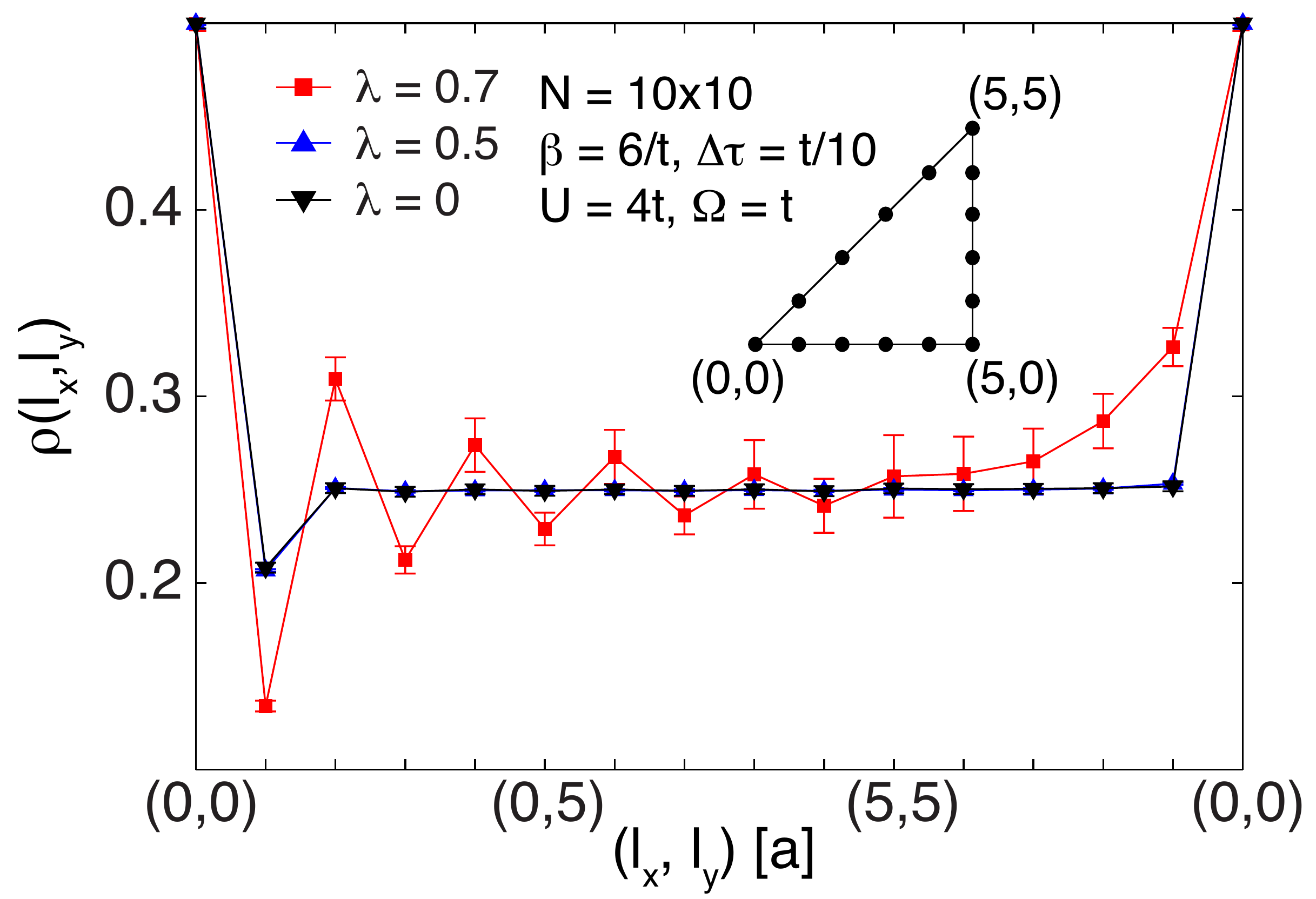}
    \caption{\label{Fig:CDW_corr}(color online) 
    The real space structure 
    of the density-density correlation function along the path indicated 
    in the inset.  For $\lambda = 0$ (black $\triangledown$) 
    the density of the system is uniform within error bars. 
    This uniform density persists for increasing values of $\lambda \le 0.5$  
    (blue $\triangle$).  However, for $\lambda = 0.7 > U/W$ (red $\square$) 
    a $(\pi,\pi)$ charge-density-wave correlation begins to develop.}
\end{figure}

In Fig. \ref{Fig:Spipi} we plot the structure factor $S(\pi,\pi)$ at 
the antiferromagnetic ordering vector for a series of half-filled clusters 
with $U = 4t$.  The data are plotted as a function of inverse temperature 
and for various values of the \eph coupling strength, as indicated in the figure.   
The $\lambda = 0$ results well reproduce the results of White {\it et al.} 
\cite{WhitePRB1989} for the Hubbard model.  However, the 
suppression of the AFM correlations as a function of $\lambda$ is apparent 
and $S(\pi,\pi)$ is reduced over the entire temperature range 
for finite values of $\lambda$.   
The suppression of the AFM order  
is also evident in the structure of the real space spin-spin correlation function 
$c(l_x,l_y)$, as shown in Fig. \ref{Fig:SCF}. 
The results for $\lambda = 0$ show a clear staggered moment in the real space 
spin structure. However, for $\lambda = 0.7$, which is   
below the peak in the CDW susceptibility 
(see Fig. \ref{Fig:Susceptibilities}b), the spin correlations 
resemble the result obtained in the paramagnetic metallic 
state.\cite{WhitePRB1989} 
This behavior is also reflected in the real-space density correlation 
function, shown in Fig. \ref{Fig:CDW_corr} for the same parameter set. 
For weak \eph coupling the cluster has a uniform charge 
distribution however upon increasing $\lambda$ to $0.7 > U/W$ a clear 
($\pi,\pi$) charge-density-wave forms.  
The behavior of both of these correlation functions implies the presence of an 
intervening metallic state below the onset of the CDW transition. 

\section{Energetics at half-filling}\label{Sec:energy}
In this final section we present results for the energetics 
of the lattice and electronic degrees of freedom. 
Again we restrict ourselves 
to half-filling and examine the energetics across the 
AFM/CDW transition.  
We first examine the average kinetic energy of the electrons $K_{el}$, 
which is defined as 
\begin{equation}
K_{el} = \left\langle -t \sum_{\langle i,j\rangle,\sigma} 
c^\dagger_{i,\sigma}c^{\phantom{\dagger}}_{j,\sigma}
\right\rangle. 
\end{equation} 

\begin{figure}
 \includegraphics[width=0.75\columnwidth]{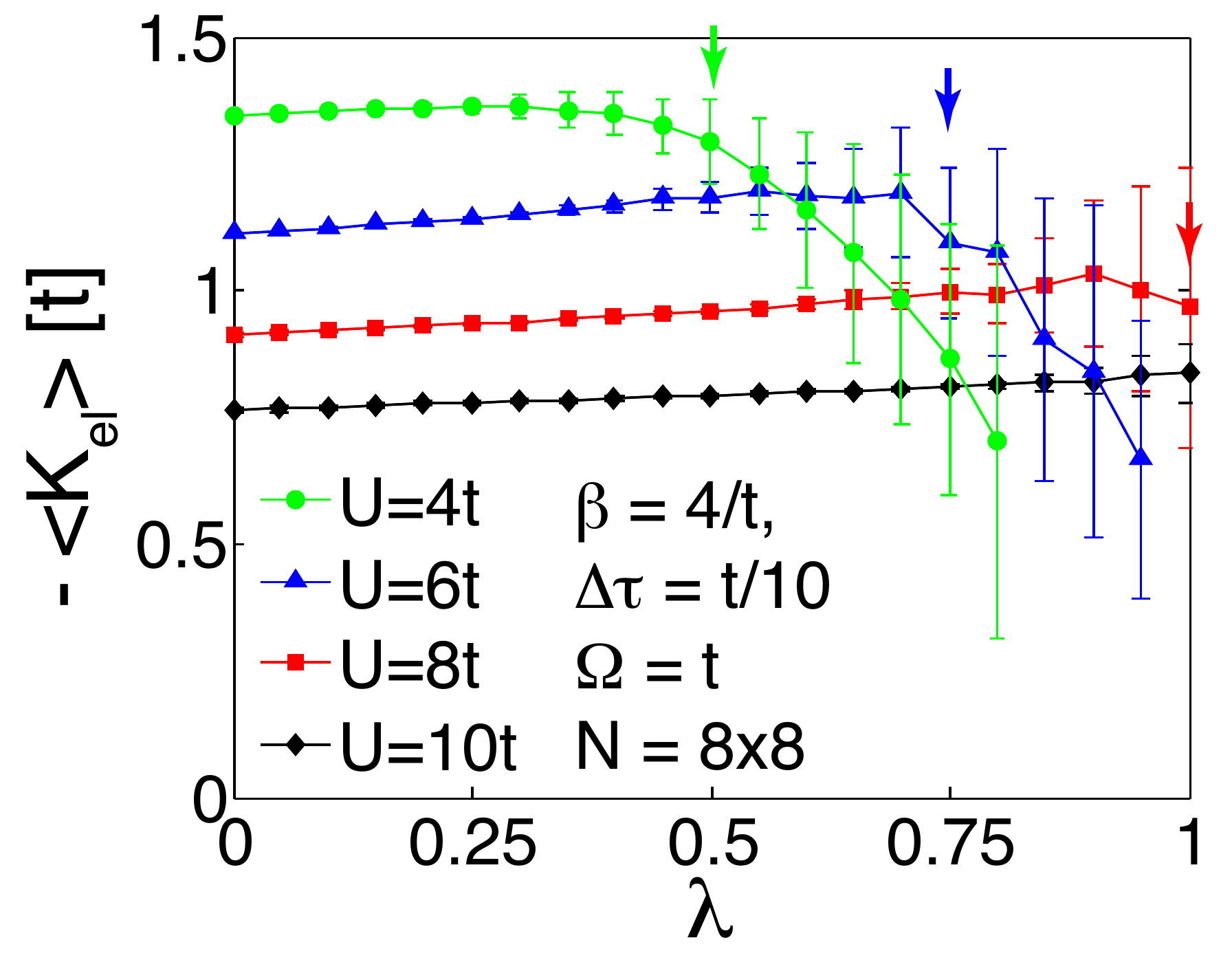}
 \caption{\label{Fig:KE_electron} (color online) 
 The negative of the average electron kinetic 
 energy as a function of the \eph interaction 
 strength $\lambda$ and $U = 4t$ (green $\bigcirc$), $6t$ (blue $\triangle$), 
 $8t$ (red $\square$), and $10t$ (black $\diamond$).  
 The arrows indicate the value of coupling when $W\lambda = U$ for 
 each data set.
 }
\end{figure}

Fig. \ref{Fig:KE_electron} shows the negative of $K_{el}$   
plotted as a function of \eph coupling 
and for values of $U$ between $4t$ and $10t$.  For $\lambda = 0$ charge fluctuations are 
suppressed by the Hubbard interaction and $-\langle K_{el}\rangle$ decreases for 
increasing values of $U$.  As $\lambda$ increases the effective Hubbard 
interaction is lowered and $K_{el}$ decreases slowly as a function of $\lambda$.  
For reference, $K_{el} \sim -1.567t$ in the non-interacting limit. 
However, once $\lambda \sim U/W$ (indicated by the arrows)  
$K_{el}$ turns over and increases rapidly. 
The value of $\lambda$ at which this occurs 
coincides with both a pronounced change in the lattice potential energy (see below) 
and the onset of the CDW susceptibility.\cite{NowadnickPRL2012}  
In Ref. \onlinecite{BauerPRB2010_2} 
similar behavior was observed in an assumed AFM ordered state. 

\begin{figure}
 \includegraphics[width=0.75\columnwidth]{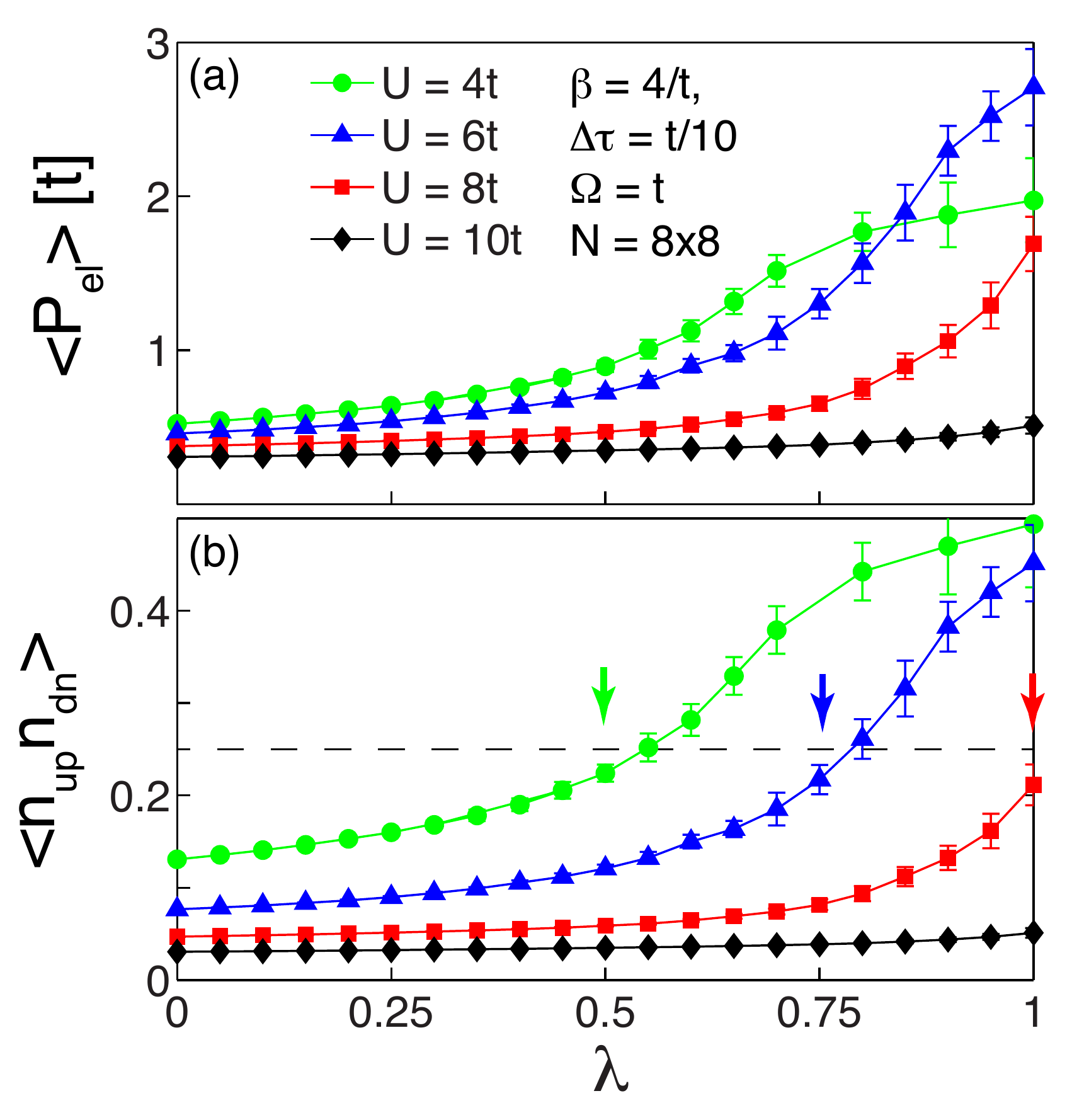}
 \caption{\label{Fig:PE_electron} (color online) (a) The average potential 
 energy of the electrons due to the Hubbard interaction $P_{el}$ as a 
 function of $\lambda$ and $U = 4t$ (green $\bigcirc$), $6t$ (blue $\triangle$),
 $8t$ (red $\square$), and $10t$ (black $\diamond$). (b) The corresponding 
 average value of the double occupancy.  The dash line indicates the value 
 expected for the non-interacting metallic system. 
 The arrows indicate the value of coupling when $W\lambda = U$ for 
 each data set.}
\end{figure}

The average potential energy of the electrons, which is proportional to the 
average number of doubly-occupied sites 
\begin{equation}
\langle P_{el} \rangle= \left\langle \sum_{i} U\hat{n}_{i,\uparrow}\hat{n}_{i,\downarrow} 
\right\rangle,
\end{equation}
is plotted in Fig. \ref{Fig:PE_electron}a.  The average value  
of the double occupancy $\langle n_{\uparrow}n_{\downarrow} \rangle$ 
appears in Fig. \ref{Fig:PE_electron}b for reference. 
(The value for a non-interacting system is indicated by the 
dashed line.) Again one sees 
the apparent competition between the AFM and CDW orders.  
For $\lambda = 0$ the system is dominated by the Hubbard interaction 
and the number of double-occupied sites is low and $P_{el}$ is lowered  
and for increasing $U$. 
When the \eph coupling increases  
$\langle n_\uparrow n_\downarrow \rangle$ grows.  This happens slowly 
at small values of $\lambda$.  However, once $\lambda \sim U/W$ the number of 
doubly-occupied sites grows more rapidly before saturating at a value 
of $0.5$ where half of the sites are doubly  
occupied as expected for $\bq = (\pi,\pi)$ CDW order. Similarly, 
the electronic potential energy increases concomitantly with the 
increase in the cost 
of this double occupancy.  This large cost in $P_{el}$ is compensated for 
by the gain in energy associated with the \eph interaction (see below).   

The behavior of $\langle n_{\uparrow}n_{\downarrow} \rangle$ shown in 
Fig. \ref{Fig:PE_electron} shows some differences from the results of 
infinite dimension 
DMFT.\cite{BauerPRB2010_2} Generically we see the growth in 
double occupancy occurring much more gradually than the DMFT result 
for the largest values of $U$.  
This appears to be the case 
regardless of the underlying state (charge ordered or normal) assumed in 
the DMFT calculations.  One possible source for this difference is the  
presence of 
the intervening metallic state in two dimensions. If such a state were present 
one would expect to see $\langle n_{\uparrow}n_{\downarrow}\rangle$ flatten 
at $1/4$   
as a function of $\lambda$ in this parameter regime.  The thermal fluctuations 
present in our calculation would then broaden this to produce milder behavor like 
that shown here.  

The average values of the phonon kinetic and potential energies are 
given by 
\begin{eqnarray}
 \langle P_{ph} \rangle&=&\frac{M\Omega^2}{2}\left\langle \sum_{i,l} X^2_{i,l} \right\rangle \\ 
 \langle K_{ph} \rangle&=&\frac{1}{2\Delta\tau} - \frac{M}{2}
 \left\langle \sum_{i,l}\bigg( \frac{X_{i,l+1}-X_{i,l}}{\Delta\tau} \bigg)^2\right\rangle. 
\end{eqnarray}
The factor of $1/(2\Delta\tau)$ appearing in the kinetic energy 
term is a Euclidean correction introduced by the Wick rotation to the 
imaginary time axis.
In the case of the lattice 
potential energy, we have subtracted off the contribution associated with the 
shift in the lattice equilibrium position in order to obtain 
a measure of the lattice fluctuations about equilibrium.  

\begin{figure}
 \includegraphics[width=0.75\columnwidth]{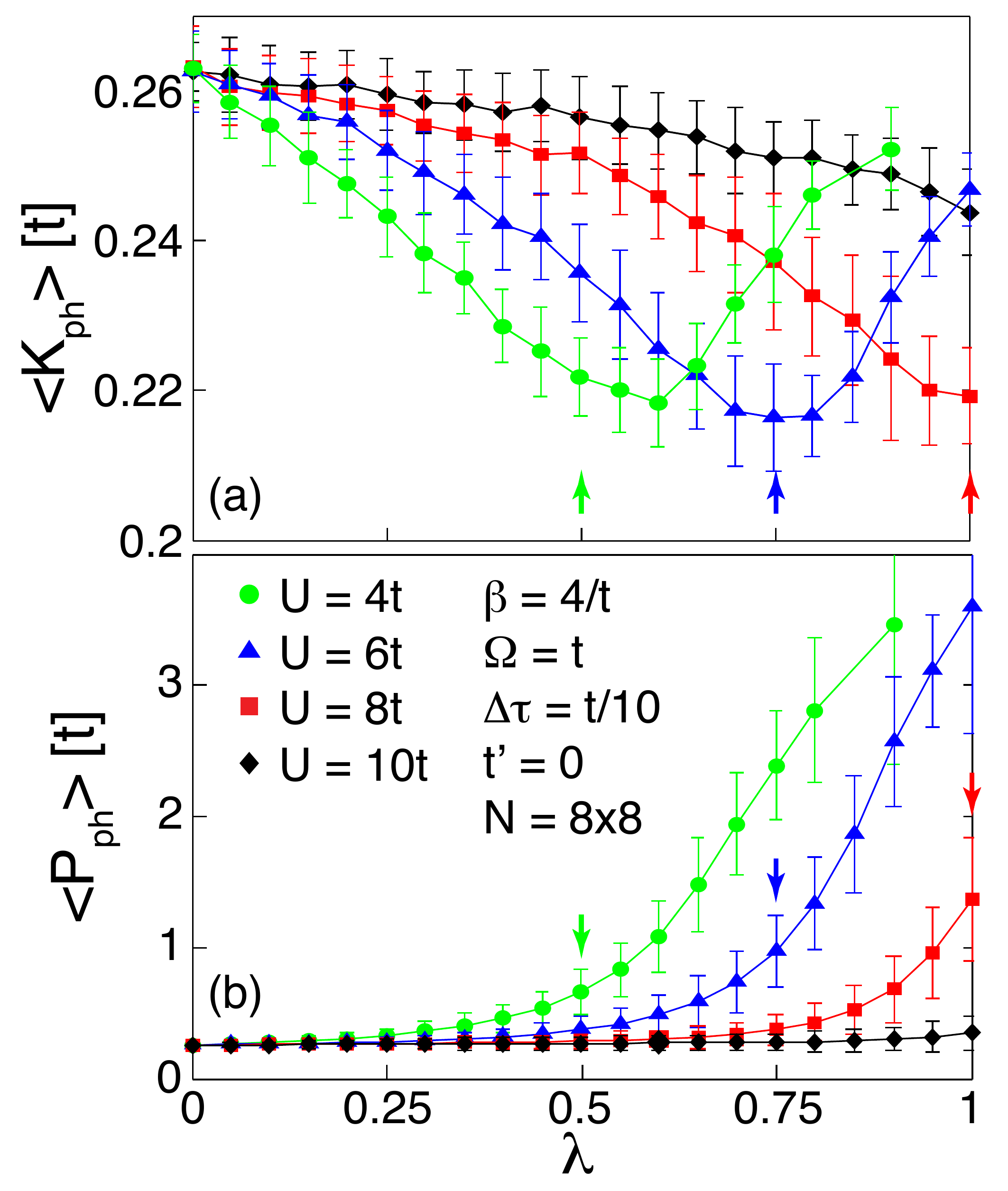}
 \caption{\label{Fig:E_phonons} (color online) 
 The average (a) kinetic and (b) potential energy of the 
 lattice for the Hubbard-Holstein model as a function of the \eph interaction 
 strength $\lambda$ and $U = 4t$ (green $\bigcirc$), $6t$ (blue $\triangle$), 
 $8t$ (red $\square$), and $10t$ (black $\diamond$).
 }
\end{figure}

The average values of the phonon kinetic and potential energies are 
shown in Figs. \ref{Fig:E_phonons}a and \ref{Fig:E_phonons}b, respectively, 
as a function of $\lambda$ and $U$.  
For $\lambda = 0$ we 
recover the atomic result $\langle K_{ph} \rangle = \langle P_{ph} \rangle = 
\frac{\Omega}{2}[n_b(\omega)-1/2]$, where $n_b(\omega) = [\exp(\omega\beta)-1]^{-1}$ 
is the bose occupation number. 
For finite \eph coupling, the kinetic (potential) energy of 
the lattice slowly decreases (increases) for $\lambda \le U/W$.  
This reflects a small renormalization of the phonons by scattering processes.
A further increase in $\lambda$ crosses the transition point 
 at which point the kinetic energy reaches a minimum before returning 
to a value comparable to that at $\lambda = 0$ with a 
concomitant increase in the potential energy.  
Again, the minimum in $K_{ph}$ and onset in the $P_{ph}$ coincide 
with the peak in the CDW susceptibilities reported in Fig. 1b of 
Ref. \onlinecite{NowadnickPRL2012}. Therefore these changes are  
linked to the onset of the CDW correlations and  
lattice's checkerboard displacement pattern. 

\begin{figure}
 \includegraphics[width=0.75\columnwidth]{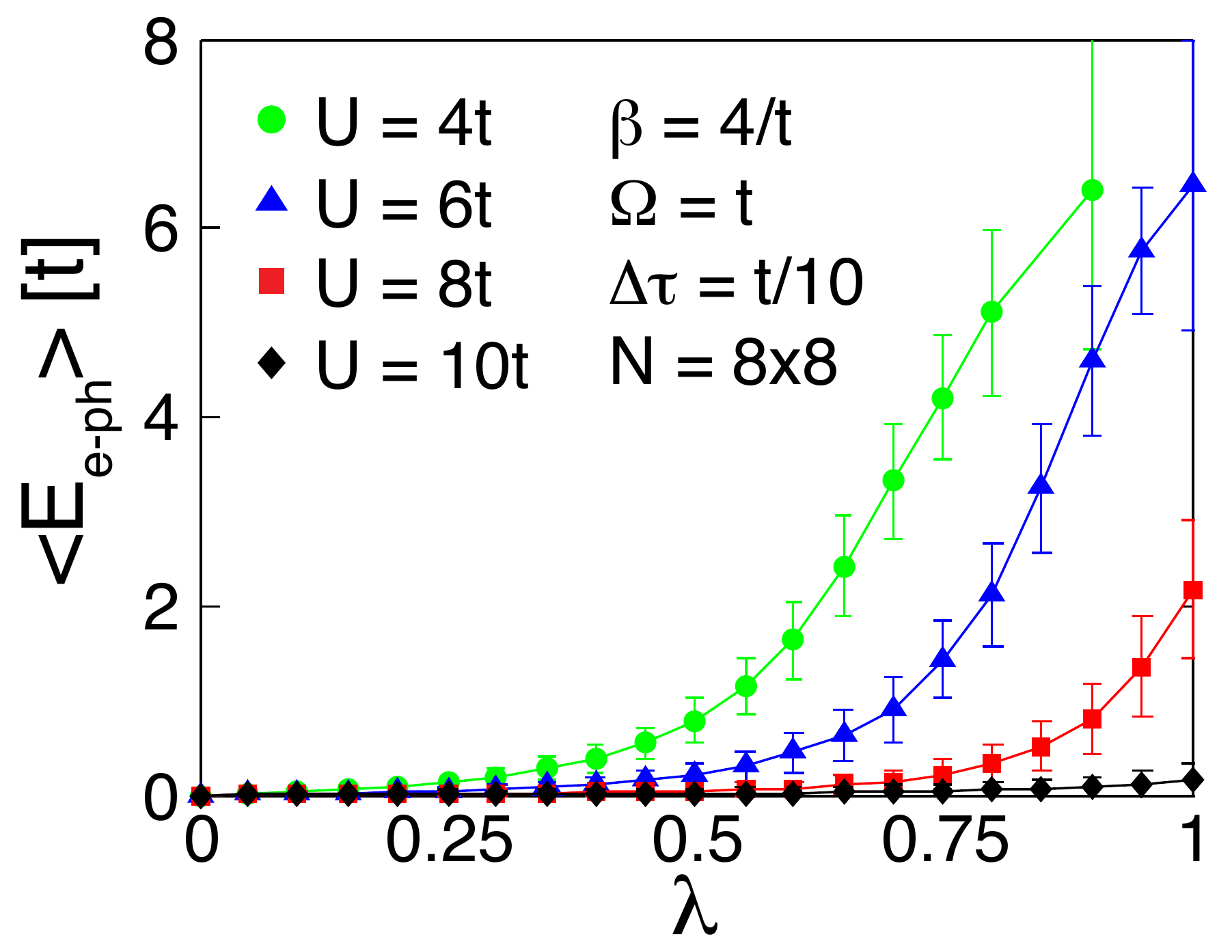}
 \caption{\label{Fig:E_elph} (color online)
 The average \eph interaction energy as a function of the \eph interaction 
 strength $\lambda$ and $U = 4t$ (green $\bigcirc$), $6t$ (blue $\triangle$), 
 $8t$ (red $\square$), and $10t$ (black $\diamond$).
 }
\end{figure}

The total phonon energy is dominated by $P_{ph}$ and therefore the 
onset of the CDW correlations is marked by an accompanying increase in the 
electronic and lattice potential energy, consistent with 
the DMFT results in infinite dimensions. This is perhaps expected as the 
CDW state is associated with an increase in double occupied sites as well 
as large lattice distortions in the checkerboard arrangement.  As previously 
mentioned, this energy comes from a corresponding gain in the \eph energy  
$E_{e-ph} = -\langle \sum_i gn_iX_i \rangle$ as shown in Fig. \ref{Fig:E_elph}.  
As with the phonon potential energy, $E_{e-ph}$ shows a weak dependence 
for $\lambda < U/W$ which gives way to a rapid rise at the onset point 
of the CDW correlations.   

\section{Concluding Remarks} \label{Sec:Discussion}
We have presented the DQMC method applied to the two-dimensional HH 
model.  In extending the DQMC algorithm to include lattice degrees of freedom 
we have found that care must be paid to the manner in which the phonon fields are 
sampled in order to ensure that one obtains the proper non-interacting limits. 
Once implemented, we benchmarked the algorithm and examined the severity of the 
fermion sign problem.  Here we found that although the phonons introduce a 
sign problem where it was originally protected by particle-hole symmetry, they do 
not significantly change the value at finite carrier concentrations where DQMC 
typically performs poorly.  This leaves open the possibility of examining 
carrier concentrations relevant to the high-T$_c$ cuprates, which we leave for 
future work.  We also found that the degree of retardation 
had a strong influence on the severity of the induced sign problem. 
However, we also observed a recovery of the fermion sign when $\lambda W \ge U$ 
and CDW correlations dominate. This suggests that parameter regimes corresponding 
to strongly correlated polarons may be accessible to DQMC. 

Focusing on the half-filled model, we also presented further evidence 
for competition between the AFM and CDW ordered phases driven by the Hubbard 
and Holstein interactions, respectively. 
This work complements our previous findings,\cite{NowadnickPRL2012} and we 
see clear, systematic suppression of the AFM correlations as $\lambda$  
increases.  In all our metrics we found that for $\lambda W \sim U$ various 
quantities appear to be similar to the values one might expect for a metallic 
phase, providing further evidence in support of the presence of an intervening 
metallic phase between 
the CDW and AFM states, at least at high temperatures.  Our results also indicate 
the importance of treating both interactions on equal footings.  In the DQMC 
treatment, the \eph interaction is capable of destabilizing the AFM correlations 
and thus addressing true competition.  This is not true for $t$-$J$-Holstein 
model treatments where a robust AFM persists for all values of $\lambda$.  
Thus one would like to revisit the issue of polaron formation using 
methods like the one presented here.   


\section{Acknowledgements} S. J. and E. A. N. contributed equally to this work.
We thank N. Nagaosa, A. S. Mishchenko, and N. Bl{\" u}mer for useful discussions. We acknowledge
support from the U. S. Department of Energy, Office of Basic Energy Sciences,
Materials Science and Engineering Division under Contract Numbers
DE-AC02-76SF00515 and DE-FC0206ER25793. 
The work of R.T.S. was supported by the National Nuclear Security Administration 
under the Stewardship Science Academic Alliances program through DOE 
Research Grant \#DE-NA0001842-0. 
S.J. acknowledges support from NSERC and SHARCNET (Canada). 
E.A.N. acknowledges support from the East Asia and Pacific Summer 
Institutes for U.S. Graduate Studies.  
Y.F.K. was supported by the Department of Defense (DoD) through 
the National Defense Science and Engineering Graduate 
Fellowship (NDSEG) Program and by the National Science 
Foundation Graduate Research Fellowship under Grant No. 1147470.  
The computational work was made possible in part by the facilities of 
SHARCNET and Compute Canada as well as the
National Energy Research Scientific Computing Center (NERSC), which is 
supported by the Office of Science of the US Department of Energy under 
contract no. DE-AC02-05CH11231. 

\appendix
\section{Average Lattice Displacement}\label{Sec:lattice}
\begin{figure}
 \includegraphics[width=0.8\columnwidth]{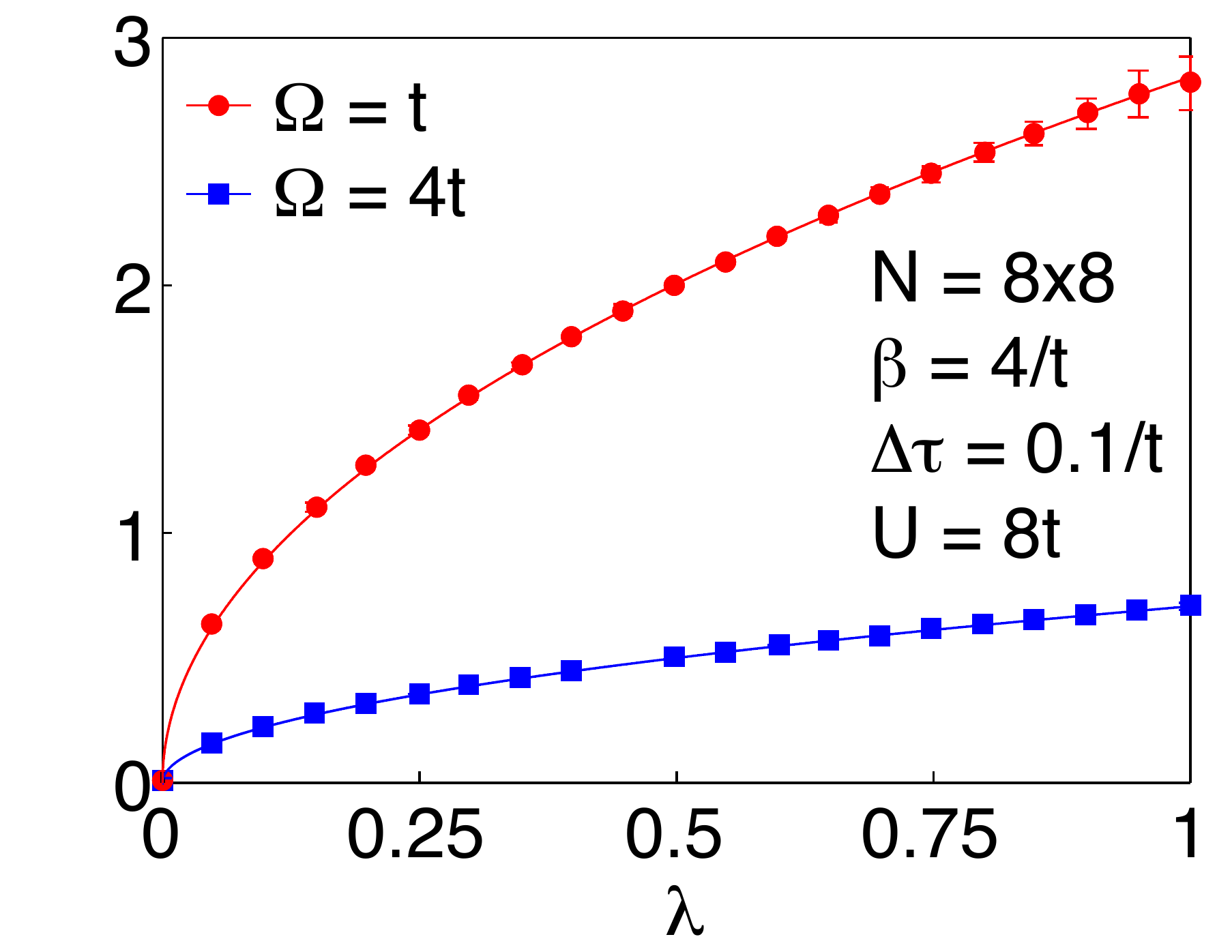}
 \caption{\label{Fig:AvgX} (Color online) The average value of the lattice 
 displacement $\langle X\rangle$ for the half-filled model as a function 
 of \ep coupling $\lambda$.  Results are shown for $\Omega = t$ (red, $\circ$) 
 and $\Omega = 4t$ (blue $\square$). The remaining parameters are as indicated. 
 The solid lines are of the form $\langle X\rangle = \sqrt{W\lambda}/\Omega$.} 
\end{figure}

On warm-up the average value of the 
lattice position $X_{i,l}$ shifts to a non-zero equilibrium position. This is 
the result of the coupled system minimizing its energy by exploiting the 
\eph interaction energy at the expense of the lattice potential energy paid for the 
shifted equilibrium position.  For a uniform charge density which one would 
expect for the half-filled case dominated by the Hubbard interaction, 
this lattice shift can be obtained 
by minimizing the total energy with respect to the phonon position.   
The new equilibrium position is given by 
\begin{equation}
\frac{d}{dX}\left[ \frac{M\Omega^2}{2}X^2 - g\langle n\rangle X\right] = 0
\end{equation}
which for $\langle n \rangle = 1$ yields $X = g/M\Omega^2 = \sqrt{W\lambda}/\Omega$. 
In Fig. \ref{Fig:AvgX} we plot $\langle X\rangle$ as a function of $\lambda$ for 
$\Omega = t$ and $4t$. The data are well fit by the functional 
form $\langle X\rangle = \sqrt{W\lambda}/\Omega$
shown as the solid lines in the plot.  This demonstrates that at half-filling 
the lattice shifts to a new equilibrium position and electrons couple to  
fluctuations around this point. This 
shift also accounts for the functional form of the renormalized chemical 
potential shift $\mu = -W\lambda$ used in Figs. \ref{Fig:filling} and 
\ref{Fig:Compressibility}.  

In general we have found that the DQMC algorithm begins to encounter 
numerical instabilities for phonon frequencies well into the adiabatic limit. 
The shift in equilibrium position is one of the possible sources for this 
instability - as the 
average lattice displacement gets large numerical overflows in 
the multiplication of the $B$ matrices begin to occur due to the exponential dependence in $X_{i,l}$.  
This difficulty could be overcome by writing the interaction term in the form 
$\sum_{i,\sigma} g(n_{i,\sigma} - \langle n \rangle)X_i$ provided the expectation 
value of the filling is known and the charge density is uniform. 
At half-filling such a procedure would be easy to implement however for finite doping 
a self-consistency loop would have to be built into the warm-up procedure. 
Furthermore, this procedure would likely do little to help in the CDW ordered phases 
once the average filling per site alternates from zero and two.


\begin{thebibliography}{99}
\bibitem{BCS}
B. Bardeen, J. Copper, and J. R. Schrieffer, Phys. Rev. {\bf 108}, 1175 (1957). 
\bibitem{Migdal1}
D. J. Scalapino in {\it Superconductivity}, edited by R. D. 
Parks (Dekker, New York, 1969), Vol. 1.
\bibitem{GrunerRMP1988}
G. Gr{\"u}ner, Rev. Mod. Phys. {\bf 60}, 1129 (1988). 
\bibitem{Migdal2}
A. B. Migdal, Soviet Physics JETP {\bf 34}, 996 (1958).
\bibitem{Eliashberg}
G. M. Eliashberg, JETP {\bf 11}, 696 (1960). 
\bibitem{ShenPRL2004}
K. M. Shen, F. Ronning, D. H. Lu, W. S. Lee, N. J. C.
Ingle, W. Meevasana, F. Baumberger, A. Damascelli, N. P.
Armitage, L. L. Miller, et al., Phys. Rev. Lett. {\bf 93}, 267002
(2004);
K. M. Shen, F. Ronning, W. Meevasana, D. H. Lu, N. J. C.
Ingle, F. Baumberger, W. S. Lee, L. L. Miller, Y. Kohsaka,
M. Azuma, et al., Phys. Rev. B {\bf 75}, 075115 (2007).
\bibitem{BoncaPRB2008}
J. Bon{\v c}a, S. Maekawa, T. Tohyama, and P. Prelov{\v s}ek, 
Phys. Rev. B {\bf 77}, 054519 (2008). 
\bibitem{CataudellaPRL2007}
V. Cataudella, G. De Filippis, A. S. Mishchenko, and N. Nagaosa, 
Phys. Rev. Lett. {\bf 99}, 226402 (2007).
\bibitem{MishchenkoPRL2004}
A. S. Mishchenko and N. Nagaosa, Phys. Rev. Lett. {\bf 93}, 036402 (2004). 
\bibitem{MishchenkoPRL2008}
A. S. Mishchenko, N. Nagaosa, Z.-X. Shen, G. De Filippis,
V. Cataudella, T. P. Devereaux, C. Bernhard, K. W. Kim,
and J. Zaanen, Phys. Rev. Lett. {\bf 100}, 166401 (2008).
\bibitem{DeFilippisPRB2009}
G. De Filippis, V. Cataudella, A. S. Mishchenko, C. A. Perroni, and N. Nagaosa, 
Phys. Rev. B {\bf 80}, 195104 (2009). 
\bibitem{CukPRL2004}
T. Cuk, F. Baumberger, D. H. Lu, N. Ingle, X. J. Zhou, 
H. Eisaki, N. Kaneko, Z. Hussain, T. P. Devereaux, N. Nagaosa, and Z.-X. Shen, 
Phys. Rev. Lett. {\bf 93}, 117003 (2004); 
T. P. Devereaux, T. Cuk, Z.-X. Shen, and N. Nagaosa, 
Phys. Rev. Lett. {\bf 93}, 117004 (2004).
\bibitem{LanzaraNature2001}
A. Lanzara, P. V. Bogdanov, X. J. Zhou, S. A. Kellar, D. L. Feng, 
E. D. Lu, T. Yoshida, H. Eisaki, A. Fujimori, K. Kishio, 
J. -I. Shimoyama, T. Noda, S. Uchida, Z. Hussain and Z. X. Shen. 
Nature {\bf 412}, 510 (2001).
\bibitem{KordyukPRL2006}
A. A. Kordyuk, S. V. Borisenko, V. B. Zabolotnyy, J. Geck, 
M. Knupfer, J. Fink, B. B{\"u}chner, C. T. Lin, B. Keimer, 
H. Berger, A. V. Pan, Seiki Komiya, and Yoichi Ando, 
Phys. Rev. Lett. {\bf 97}, 017002 (2006). 
\bibitem{JohnsonPRL2001}
P. D. Johnson, T. Valla, A. V. Fedorov, Z. Yusof, B. O. Wells, 
Q. Li, A. R. Moodenbaugh, G. D. Gu, N. Koshizuka, C. Kendziora, 
Sha Jian, and D. G. Hinks, Phys. Rev. Lett. {\bf 87}, 177007 (2001).  
\bibitem{DahmNaturePhysic2009}
T. Dahm, V. Hinkov, S. V. Borisenko, A. A. Kordyuk, V. B. Zabolotnyy, 
J. Fink, B. B{\"u}chner, D. J. Scalapino, W. Hanke, and B. Keimer, 
Nature Phys. {\bf 5}, 217 (2009).
\bibitem{JohnstonACMP2010}
S. Johnston, W. S. Lee, Y. Chen, E. A. Nowadnick, B. Moritz, Z.-X. Shun, 
and T. P. Devereaux, Advances in Condensed Matter Physics {\bf 2010}, 
968304 (2010).
\bibitem{PlumbPRL2010}
N. C. Plumb, T. J. Reber, J. D. Koralek, Z. Sun, J. F. Douglas, 
Y. Aiura, K. Oka, H. Eisaki, and D. S. Dessau, Phys. Rev. Lett. {\bf 105}, 
046402 (2010). 
\bibitem{VishikPRL2012}
I. M. Vishik, W. S. Lee, F. Schmitt, B. Moritz, T. Sasagawa, 
S. Uchida, K. Fujita, S. Ishida, C. Zhang, T. P. Devereaux, 
and Z. X. Shen, Phys. Rev. Lett. {\bf 104}, 207002 (2010).  
\bibitem{LeePRB2008}
W. S. Lee, W. Meevasana, S. Johnston, D. H. Lu, I. M. Vishik, 
R. G. Moore, H. Eisaki, N. Kaneko, T. P. Devereaux, and Z. X. Shen, 
Phys. Rev. B {\bf 77}, 140504 (2008). 
\bibitem{AnzaiPRL2010}
H. Anzai, A. Ino, T. Kamo, T. Fujita, M. Arita, H. Namatame, 
M. Taniguchi, A. Fujimori, Z.-X. Shen, M. Ishikado, and S. Uchida, 
Phys. Rev. Lett. {\bf 105}, 227002 (2010).
\bibitem{RameauPRB2009}
J. D. Rameau, H.-B. Yang, G. D. Gu, and P. D. Johnson, Phys. Rev. B 
{\bf 80}, 184513 (2009).
\bibitem{MeevasanaPRL2006}
W. Meevasana, N. J. C. Ingle, D. H. Lu, J. R. Shi, F. Baumberger, 
K. M. Shen, W. S. Lee, T. Cuk, H. Eisaki, T. P. Devereaux, 
N. Nagaosa, J. Zaanen, and Z.-X. Shen, Phys. Rev. Lett. {\bf 96}, 157003 (2006). 
\bibitem{LeeNature2006} 
J. Lee, K. Fujita, K. McElroy, J. A. Slezak, M. Wang,
Y. Aiura, H. Bando, M. Ishikado, T. Masui, J.-X. Zhu,
{\it et al.}, Nature (London) {\bf 442}, 546 (2006)
\bibitem{Pasupathy}
A. N. Pasupathy, A. Pushp, K. K. Gomes, C. V. Parker, J. Wen, Z. Xu, 
G. Gu, S. Ono, Y. Ando, and A. Yazdani, Science {\bf 320}, 196 (2008). 
\bibitem{JenkinsPRL2009}
N. Jenkins, Y. Fasano, C. Berthod, I. Maggio-Aprile, 
A. Piriou, E. Giannini, B. W. Hoogenboom, C. Hess, T. Cren, and 
{\O}. Fischer, Phys. Rev. Lett. {\bf 103}, 227001 (􏰀2009).
\bibitem{ZasadzinskiPRL2006}
J. F. Zasadzinski, L. Ozyuzer, L. Coffey, K. E. Gray, 
D. G. Hinks, and C. Kendziora, Phys. Rev. Lett. {\bf 96}, 017004 (2006). 
\bibitem{deCastroPRL2008}
G. Levy de Castro, C. Berthod, A. Piriou, E. Giannini, and 
{\O}. Fischer, Phys. Rev. Lett. {\bf 101}, 267004 (2008). 
\bibitem{ZhuPRB2006}
J.-X. Zhu, A. V. Balatsky, T. P. Devereaux, Q. Si, J. Lee, K. McElroy, 
and J. C. Davis, Phys. Rev. B {\bf 73}, 014511 (2006); 
J.-X. Zhu, K. McElroy, J. Lee, T. P. Devereaux, Q. Si, J. C. Davis, 
and A. V. Balatsky, Phys. Rev. Lett {\bf 97}, 177001 (2006).  
\bibitem{JohnstonDOS}
S. Johnston and T. P. Devereaux, Phys. Rev. B {\bf 81}, 214512 (2010).
\bibitem{GuomengZhao}
Guo-meng Zhao, Phys. Rev. B {\bf 75}, 214507 (2007); 
Guo-meng Zhao, Phys. Rev. Lett. {\bf 103}, 236403 (2009).
\bibitem{CarbotteReview}
J. P. Carbotte, T. Timusk, and J. Hwang, 
Reports on Progress in Physics {\bf 74}, 066501 (2011).
\bibitem{EvH}
E. van Heumen, E. Muhlethaler, A. B. Kuzmenko, H. Eisaki, 
W. Meevasana, M. Greven and D. van der
Marel, Phys. Rev. B {\bf 79}, 184512 (2009).
\bibitem{LeePreprint}
W. S. Lee, S. Johnston, B. Moritz, J. Lee, M. Yi, K. J. Zhou, 
T. Schmitt, L. Patthey, V. Strocov, K. Kudo, Y. Koike, 
J. van den Brink, T. P. Devereaux, and Z. X. Shen, 
arXiv:1301.4267 (2013). 
\bibitem{MillisNature1998}
A. J. Millis, Nature (London) {\bf 392}, 147 (1998). 
\bibitem{MillisPRB1996}
A. J. Millis, R. Mueller, and B. I. Shraiman, Phys. Rev. B 
{\bf 54}, 5405 (1996). 
\bibitem{MannellaPRB2007}
N. Mannella, W. L. Yang, K. Tanaka, X. J. Zhou, H. Zheng, 
J. F. Mitchell, J. Zaanen, T. P. Devereaux, N. Nagaosa, Z. Hussain, 
and Z.-X. Shen, Phys. Rev. B {\bf 76}, 233102 (2007).
\bibitem{Durand2003}
P. Durand, G. R. Darling, Y. Dubitsky, A. Zaopo, and M. J. Rosseinsky, 
Nature Materials {\bf 2}, 026401 (2003). 
\bibitem{CaponeScience2002}
M. Capone, M. Fabrizio, C. Castellani, and E. Tosatti, 
Science {\bf 296}, 2364 (2002); 
M. Capone, M. Fabrizio, C. Castellani, and E. Tosatti, 
Rev. Mod. Phys. {\bf 81}, 943 (2009). 
\bibitem{GunnarssonRMP1997}
O. Gunnarsson, Rev. Mod. Phys. {\bf 69}, 575 (1997). 
\bibitem{HanPRL2003}
J. E. Han, O. Gunnarsson, and V. H. Crespi, Phys. Rev. Lett. {\bf 90}, 
167006 (2003).
\bibitem{MedardeJPCM1997}
M. L. Medarde, J. Phys.: Condens. Matter {\bf 9}, 1679 (1997). 
\bibitem{LauPreprint}
B. Lau and A. J. Millis, arXiv:1210.6693 (2012).
\bibitem{AlexandrovPRL2011}
A. S. Alexandrov and V. V. Kabanov, Phys. Rev. Lett. {\bf 106}, 136403 (2011).
\bibitem{AlexandrovPRB1996}
A. S. Alexandrov, Phys. Rev. B {\bf 53}, 2863 (1996).
\bibitem{MeevasanaScreening}
W. Meevasana, T. P. Devereaux, N. Nagaosa, Z.-X. Shen,
and J. Zaanen, Phys. Rev. B {\bf 74}, 174524 (2006).
\bibitem{JohnstonPRL2012}
S. Johnston, I. M. Vishik, W. S. Lee, F. Schmitt, S. Uchida, 
K. Fujita, S. Ishida, N. Nagaosa, Z. X. Shen, and 
T. P. Devereaux, Phys. Rev. Lett. {\bf 108}, 166404 (2012). 
\bibitem{JohnstonPRB2010}
S. Johnston, F. Vernay, B. Moritz, Z.-X. Shen, N. Nagaosa, 
J. Zaanen, and T. P. Devereaux, Phys. Rev. B {\bf 82}, 064513 (2010). 
\bibitem{BulutPRB1996}
N. Bulut and D. J. Scalapino, Phys. Rev. B {\bf 54}, 14971 (1996).
\bibitem{MaksimovPRB2005}
E. G. Maksimov, O. V. Dolgov, and M. L. Kuli{\'c}, Phys. Rev. B {\bf 72}, 212505 (2005).
\bibitem{HuangPRB2003}
Z. B. Huang, W. Hanke, E. Arrigoni, and D. J. Scalapino, 
Phys. Rev. B {\bf 68}, 220507(R) (2003). 
\bibitem{KulicPRB1994}
M. L. Kuli{\'c} and R. Zeyher, Phys. Rev. B {\bf 49}, 4395 (1994).
\bibitem{ZeyherPRB1996}
R. Zeyher and M. L. Kuli{\'c}, Phys. Rev. B {\bf 53}, 2850 (1996).
\bibitem{BauerPRB2010}
J. Bauer and G. Sangiovanni, Phys. Rev. B {\bf 82}, 184535 (2010).
\bibitem{Pintschovius}
L. Pintschovius, Phys. Stat. Sol. (b) {\bf 242}, 30 (2005); 
M. d’Astuto, G. Dhalenne, J. Graf, M. Hoesch, P. Giura, M. 
Krisch, P. Berthet, A. Lanzara, and A. Shukla, Phys. Rev. B 
{\bf 78}, 140511(R) (2008).
\bibitem{ReznikNature}
D. Reznik, G. Sangiovanni, O. Gunnarsson, and T. P. Devereaux, 
Nature {\bf 455}, E6 (2008). 
\bibitem{DFT}
K.-P. Bohnen, R. Heid, and M. Krauss, EPL {\bf 64}, 104 (2003); 
F. Giustino, M. L. Cohen, S.-G. Louie, Nature {\bf 452}, 975 (2008).
\bibitem{RoschPRB2004}
O. R{\"o}sch and O. Gunnarsson, Phys. Rev. B {\bf 70}, 224518 (2004). 
\bibitem{HorschPhysicaB2005}
P. Horsch and G. Khaliullin, Physica B {\bf 359}, 620 (2005). 
\bibitem{MacridinPRL2006}
A. Macridin, B. Moritz, M. Jarrell, and T. Maier, Phys. Rev. Lett. {\bf 97}, 
056402 (2006); 
A. Macridin, B. Moritz, M. Jarrell, and T. Maier, J. Phys.:
Conden Mat {\bf 24}, 475603 (2012).
\bibitem{SangiovanniPRL2005}
G. Sangiovanni, M. Capone, C. Castellani, and M. Grilli,
Phys. Rev. Lett. {\bf 94}, 026401 (2005).
\bibitem{SangiovanniPRL2006}
G. Sangiovanni, O. Gunnarsson, E. Koch, C. Castellani,
and M. Capone, Phys. Rev. Lett. {\bf 97}, 046404 (2006).
\bibitem{RoschPRL2004}
O. R{\"o}sch and O. Gunnarsson, Phys. Rev. Lett. {\bf 92}, 146403 (2004). 
\bibitem{Prelovsek}
P. Prelov{\v s}ek, R. Zeyher, and P. Horsch, Phys. Rev. Lett. {\bf 96}, 
086402 (2006). 
\bibitem{BerciuPRB2007}
M. Berciu, Phys. Rev. B {\bf 75}, 081101 (2007). 
\bibitem{WhitePRB1989}
S. R. White, D. J. Scalapino, R. L. Sugar, E. Y. Loh, J. E. Gubernatis, 
and R. T. Scalettar, Phys. Rev. B {\bf 40}, 506 (1989). 
\bibitem{MarsiglioPRB1990}
F. Marsiglio, Phys. Rev. B {\bf 42}, 2416 (1990).
\bibitem{ScalettarPRB1989}
R. T. Scalettar, N. E. Bickers, and D. J. Scalapino, Phys.
Rev. B {\bf 40}, 197 (1989).
\bibitem{BauerEPL2010}
J. Bauer, Europhys. Lett. {\bf 90}, 27002 (2010).
\bibitem{BauerPRB2010_2}
J. Bauer and A. C. Hewson, Phys. Rev. B {\bf 81}, 235113 (2010).
\bibitem{ClayPRL2005}
R. T. Clay and R. P. Hardikar, Phys. Rev. Lett. {\bf 95}, 096401 (2005). 
\bibitem{FehskePRB2004}
H. Fehske, G. Wellein, G. Hager, A. Wei{\ss}e, and A. R. Bishop, 
Phys. Rev. B {\bf 69}, 165115 (2004). 
\bibitem{NowadnickPRL2012}
E. A. Nowadnick, S. Johnston, B. Moritz, R. T. Scalettar, 
and T. P. Devereaux, Phys. Rev. Lett. {\bf 109}, 246404 (2012). 
\bibitem{TrivediPRL1995}
N. Trivedi and M. Randeria, Phys. Rev. Lett. {\bf 75}, 312 (1995). 
\bibitem{BSS}
R. Blankenbecler, D. J. Scalapino, and R. L. Sugar, Phys.
Rev. D {\bf 24}, 2278 (1981).
\bibitem{HS}
    J. E. Hirsch, Phys. Rev. B {\bf 31}, 4403 (1985).
\bibitem{BergerPRB1995}
E. Berger, P. Val{\' a}{\v s}ek, and W. von der Linden, 
Phys. Rev. B {\bf 52}, 4806 (1995). 
\bibitem{Suzuki}
M. Suzuki, Prog. Theor. Phys. {\bf 56}, 1454 (1976); 
R. M. Fye, Phys. Rev. B {\bf 33}, 6271 (1986); 
R. M. Fye and R. T. Scalettar, {\it ibid}. {\bf 36}, 3833 (1987).
\bibitem{GullRMP}
E. Gull, A. J. Millis, A. I. Lichtenstein, A. N. Rubtsov, M. Troyer, 
and P. Werner, Rev. Mod. Phys. {\bf 83}, 349 (2011). 
\bibitem{ScalettarPRB1991}
A global update scheme similar in spirit to the phonon updates 
are also needed for large values 
of $U$ and $\beta$.\cite{ScalettarSub} 
In this case updates are made to multiple sites on a given time slice. 
\bibitem{ScalettarSub}
R. T. Scalettar, R. M. Noack, and R. R. P. Singh, Phys. Rev. B {\bf 44}, 
10502 (1991). 
\bibitem{auto}
One might be suspicious that the autocorrelation time for the phonon fields 
could be a factor.  To test this we performed a second simulation 
for the $\Omega = t$ where the number of measurement sweeps and spacing between 
measurements was increased by a factor of  
one hundred.  This run produced no measurable difference 
in the observed quantities.      
\bibitem{FootNote1}
We evaluate $\kappa$ by numerically differentiating a weighted 
smoothing spline fit to the $\langle n(\mu) \rangle$ data. Each data 
point is weighted by the statistical error bars shown in Fig. \ref{Fig:filling}. 
\bibitem{BauerPRB2011}
J. Bauer, J. E. Han, and O. Gunnarsson, Phys. Rev. B {\bf 84}, 184531 (2011). 
\bibitem{MA} 
M. Berciu, Phys. Rev. Lett. {\bf 98}, 209702 (2007); 
M. Berciu and G. L. Goodvin, Phys. Rev. B {\bf 76}, 165109 (2007).
\end{thebibliography}
\end{document}